\documentclass[letterpaper]{article} % DO NOT CHANGE THIS
\usepackage{aaai24}  % DO NOT CHANGE THIS
\usepackage{times}  % DO NOT CHANGE THIS
\usepackage{helvet}  % DO NOT CHANGE THIS
\usepackage{courier}  % DO NOT CHANGE THIS
\usepackage[hyphens]{url}  % DO NOT CHANGE THIS
\usepackage{graphicx} % DO NOT CHANGE THIS
\usepackage{natbib}  % DO NOT CHANGE THIS AND DO NOT ADD ANY OPTIONS TO IT
\usepackage{caption} % DO NOT CHANGE THIS AND DO NOT ADD ANY OPTIONS TO IT
\usepackage{algorithm}
\usepackage{algorithmic}

\usepackage{amsthm}
\usepackage{amsmath}
\theoremstyle{definition}
\newtheorem{exmp}{Example}[section]
\usepackage{multirow}
\usepackage{booktabs}
\usepackage{amssymb}
\usepackage{newfloat}
\usepackage{listings}
\DeclareCaptionStyle{ruled}{labelfont=normalfont,labelsep=colon,strut=off} % DO NOT CHANGE THIS
\floatstyle{ruled}
\newfloat{listing}{tb}{lst}{}
\floatname{listing}{Listing}
\usepackage{comment}
\usepackage{xcolor}
\definecolor{highlight}{RGB}{255,0,0}

\definecolor{sanadcolor}{RGB}{0,0,255}

\definecolor{nischalcolor}{RGB}{220,68,5}

\title{How Does User Behavior Evolve During Exploratory Visual Analysis?}
\author {
    Sanad Saha\equalcontrib \textsuperscript{\rm 1},
    Nischal Aryal\equalcontrib \textsuperscript{\rm 1},
    Leilani Battle\textsuperscript{\rm 2}
    Arash Termehchy\textsuperscript{\rm 1}
}
\affiliations {
    \textsuperscript{\rm 1}Oregon State University\\
    \textsuperscript{\rm 2}University of Washington\\
    sahasa@oregonstate.edu, aryaln@oregonstate.edu, leibatt@cs.washington.edu, termehca@oregonstate.edu
}
\usepackage{bibentry}
\begin{document}

\maketitle

\begin{abstract}
Exploratory visual analysis (EVA) is an essential stage of the data science pipeline, where users often lack clear analysis goals at the start and iteratively refine them as they learn more about their data.
Accurate models of users' exploration behavior are becoming increasingly vital to developing responsive and personalized tools for exploratory visual analysis. Yet we observe a discrepancy between the \emph{static} view of human exploration behavior adopted by many computational models versus the \emph{dynamic} nature of EVA. In this paper, we explore potential parallels between the evolution of users' interactions with visualization tools during data exploration and assumptions made in popular online learning techniques. Through a series of empirical analyses, we seek to answer the question: how might users' exploration behavior evolve in response to what they have learned from the data during EVA? We present our findings and discuss their implications for the future of user modeling for system design.
\end{abstract}

\section{Introduction}
%interactive data exploration, and challenges to the user
Data analysts often explore large datasets to find relevant information or discover interesting patterns in the data using \textbf{visual exploration systems (VES)}, like Tableau~\cite{Tableau}, Microsoft PowerBI~\cite{powerbi}, etc.
They interactively query the visualizations shown by the VES until they discover their desired information.
This process, known as \textbf{exploratory visual analysis (EVA)}, is iterative and complex.
% \textbf{interactive data exploration (IDE)}, is iterative and complex.
% It is particularly challenging when the user has no clear objective and delves into the data to discover intriguing observations~\cite{battle2019characterizing, liu2014effects, zgraggen2018investigating}, i.e., performs \emph{open-ended} exploration.
EVA is particularly challenging as analysts often encounter new datasets with unknown structures and content.
% encounter new datasets whose content and structure are unknown.
% Additionally, analysis tasks based on such datasets can also be vague, e.g., discovering intriguing observations.
Complexity increases as they 
often commence EVA with only vague analysis goals in mind, e.g.,
%delve into such datasets to complete analysis tasks that are often vague, e.g.,
to discover intriguing observations~\cite{battle2019characterizing, liu2014effects}.
%It forces them to explore large datasets without any clear objective and goals \cite{battle2019characterizing, liu2014effects, zgraggen2018investigating}.
% The following example will give us more insight into how
% \sanad{While it is certainly possible to conduct IDE without visualizations, it demands significant effort from the users (e.g., analyzing tabular data with SQL queries). 
% Based on existing works on data exploration from both the database and visualization community, we identify data visualizations as the dominant method for data exploration and center our focus around them in this paper.}
The following example clarifies how
users dynamically generate hypotheses and perform multiple interactions with a VES
% \textbf{visual exploration system (VES)} 
to achieve their goals.
%overcome such obstacles and achieve desired results.

\begin{figure}[!htbp]
\centering
\vspace{-3mm}
\includegraphics[scale=0.2]{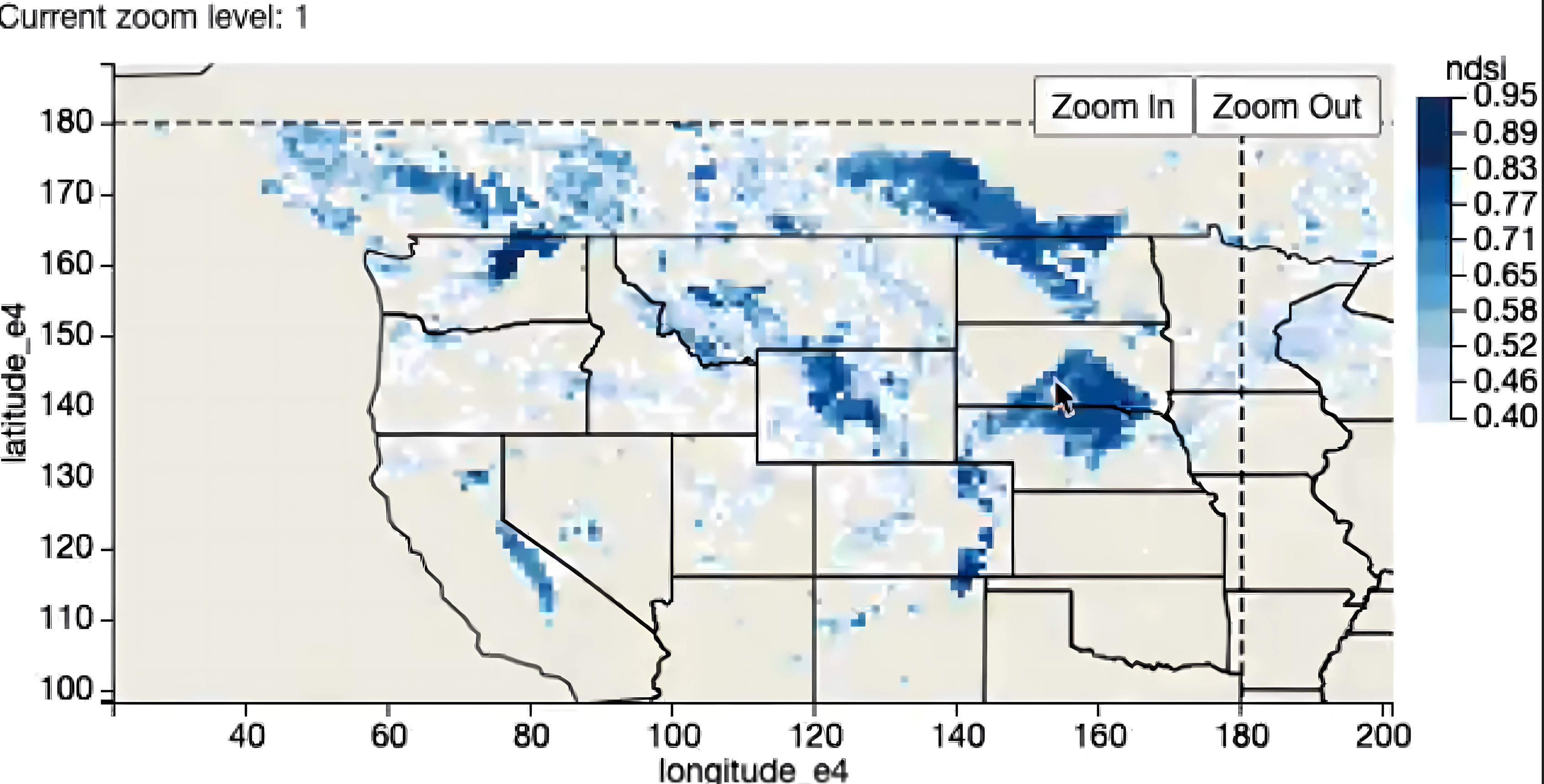}
\centering
% \Description{Map of United States showing snow levels in blue}
\vspace{-2mm}
\caption{Exploring satellite imagery using a VES.}
\vspace{-3mm}
\label{fig: forecache-interface1}
\end{figure}

\begin{exmp}
\label{example:snowlevel}
%Let us consider an analysis task where
Suppose an analyst \emph{explores snow cover from NASA satellite imagery to identify abnormally snowy regions in the USA, a potential consequence of global warming.} 
She has not analyzed this dataset before.
So, initially, she does not know what snow levels are abnormal.
She uses a VES (Figure \ref{fig: forecache-interface1}) to explore different regions using pan and zoom operations.
Through EVA, the analyst learns what areas have high snow levels, e.g., snowy mountain ranges. 
She eventually hypothesizes that outliers correspond to mountainous regions that deviate from expected high snow levels.
With this intent, she alters her interaction strategy to test her hypothesis. This process of generating and testing hypotheses continues until she gains the desired insights.  
\end{exmp}

% It is particularly challenging when the user has no clear objective and delves into the data to discover intriguing observations~\cite{battle2019characterizing, liu2014effects, zgraggen2018investigating}, i.e., performs \emph{open-ended} exploration. 

During EVA, users generate a wide range of query workloads. For instance, in \textit{Example \ref{example:snowlevel}}, the user may perform queries to compare multiple snowy regions at different levels of detail. Current systems offer optimization techniques to support demanding workloads under specific contexts, including re-using previously computed results and building specialized
data structures such as stratified samples or data cubes~\cite{battle2020structured}. But users' ad-hoc queries during EVA can thwart standard optimization techniques~\cite{battle2020database}. Studies have shown how these inefficiencies can lead to user frustration and even abandonment of exploration tasks~\cite{liu2013immens}.

%How DES aid in interactive data exploration
Understanding and modeling users' exploration behavior can help to customize VESs to match the user's interests, preferences, and exploration strategies~\cite{zeng2021evaluation}. Such systems can prefetch~\cite{battle2016dynamic}, suggest relevant or interesting data regions \cite{wongsuphasawat2016towards}, and recommend exploratory operations~\cite{milo2018next}.
%, and speed up the exploration process by prefetching promising data regions\cite{battle2016dynamic}.
In \textit{example \ref{example:snowlevel}}, a VES could infer that the user intends to compare mountainous snowy regions. Then, it can prefetch similar data areas to reduce system latency or recommend interesting regions to lessen users' exploration load. However, the VES must also detect shifts in users' exploration behavior after a hypothesis is formed, else, its predictions will become obsolete.

%what current systems get wrong on modeling user activities  
We observe that user models in current VESs implicitly assume that users' intent and strategies are \emph{fixed}, i.e., users' analysis goals and strategies for achieving it will not change much over time~\cite{battle2016dynamic}.
%need more info if this is a single paragraph
% However, due to the diverse nature of analysis tasks, dataset complexity, and structural variations, users often engage in IDE to develop a comprehensive understanding of the data. 
% They may initially start without clear intentions and dynamically generate hypotheses. 
% As they gradually accumulate knowledge, 
In contrast, users may enrich their understanding of the data during EVA, and may \textit{change} their exploration strategies to better express their intents ~\cite{battle2019characterizing}.

% However, because of the diverse analysis task goals, complexity, and structure of the datasets, user often perform multiple interactions to develop a general understanding of the data. 
% They may start without clear intents, \emph{dynamically} generate hypotheses and \textbf{learn} to refine their intents~\cite{battle2016dynamic, battle2019characterizing, liu2014effects}, i.e., \emph{changing} analysis goals and exploration strategies.
% For determining the best approach to complete an analysis task, users' goals evolve, resulting in {\it non-stationary} exploration behavior.

% Moreover, recent research using real-world keyword query workloads indicates that \emph{users learn} to improve their strategies of expressing intents by modifying their keyword queries over time\cite{cen2013reinforcement, mccamish2018data}.
% We posit that the evolutionary nature of EDA also shares strong parallels with query-based information searching.

%modeling evolution of users' exploration behavior

As users may alter their exploration behavior in response to what they learn about the data during exploration, a natural question is whether known learning methods can model users' evolving information needs in real-time. 
% for modeling these behavioral shifts.
%We propose that modeling users' data exploration behavior requires extensive empirical study.
Many online algorithms have shown promising results in modeling human behavior in game theory, cognitive psychology \cite{bush1953stochastic,niv2009reinforcement, niv2012TD}, etc. \textit{This paper addresses the absence of an empirical investigation of these human behavior algorithms as models for the evolution of user reasoning during EVA.} Identifying and evaluating existing algorithms will help us decide if and when new algorithms are required, thereby guiding future efforts in building accurate models for VESs
to further support EVA.

%Give an overview of our goal and what we do
The goal of this paper is to investigate and model the evolution of users' exploration behavior in response to user learning using influential studies from the visualization community that apply distinct approaches to capturing exploration behavior. \textit{This helps us to observe how current learning methods model behavioral changes across the gamut of exploration tasks rather than in just one tool or scenario}. In this work, we present our investigation on ~\cite{liu2014effects} and ~\cite{battle2016dynamic} for the sake of space and provide our investigation of ~\cite{battle2019characterizing} in the appendix.  
% Our primary focus in this work is to investigate users' Open-ended/Goal-directed exploration. Consequently, we've put a detailed investigation of users doing Focused/Goal-directed exploration using \citeauthor{battle2019characterizing} in the appendix. 
Specifically, we seek answers to the following questions:
% Specifically, we seek answers to the following questions through empirical analyses: 
% \squishlisttwo
% \item
(a) \emph{How does learning manifest during exploratory visual analysis?}
% \item
(b)
\emph{Does users' exploration behavior actually evolve?} and
% \item
(c)
\emph{Can existing algorithms model users' exploration behavior in EVA?}

% \squishend
% \cite{wongsuphasawat2016towards, gotz2009behavior, wongsuphasawat2015voyager}, exploratory operations \cite{milo2018next} and speeding up the process by prefetching promising data regions\cite{battle2016dynamic}. 
% By doing so, we can decrease necessary interactions and enable efficient and effective exploration \cite{battle2016dynamic}.
% Access to these models will help DES adapt to users' learning behaviors, predict future actions, and assist them in the complex process of data exploration.

%Give a list of our contribution.
% \item summary of high level findings, lessons learned section
% Our analysis is a breadth rather than depth-focused analysis. This analysis is also exploratory in nature and the first of its kind. 
% Our empirical analysis aims not to draw precise conclusions on the best approaches to modeling users' evolving behavior but to provide important hypotheses to drive future user studies to collect targeted data. 
% In the meantime, our study provides meaningful findings for the range of models tested and thus provides useful model design guidelines for developing user-adaptive VES.
% We mirror Feng et al.'s approach of leveraging prior user studies to address our research questions. 

In summary, we make the following contributions:
\vspace{-2mm}
\begin{itemize}
\item  We examine users' decision-making strategies in real-life EVA tasks with different characteristics, e.g., goals, complexity, and prior knowledge. 
\textit{Our preliminary findings demonstrate that user learning about the data affects users' exploration behavior.}
% (Section \ref{sec: considerations})
\vspace{-1mm}
\item We use statistical tests to analyze  \textit{how users' exploration strategies evolve} and connect our findings with user learning. 
Test results indicate that user exploration behavior differs depending on the clarity and complexity of the exploration task. This variance is observable across \emph{goal-directed, focused, and open-ended scenarios}.
\vspace{-1mm}
\item 
We utilize popular learning algorithms from reinforcement learning, economics, cognitive psychology, and neuroscience to model users' exploration behavior. 
% For \emph{focused tasks}, users' interactions are predicted with 75\% accuracy by these algorithms. For \emph{goal-directed tasks}, existing models struggle (68\%) and fail for \emph{open-ended} data exploration tasks (60\%).
% For \emph{focused tasks}, users' interactions are predicted with 75\% accuracy. 
They struggle more for \emph{goal-directed tasks} (68\%) and \emph{open-ended tasks} (53\%) (Subsections \ref{sec: forecache_results},\ref{sec: imMens_results}).
\vspace{-1mm}
\item
We present common findings and challenges from our investigations and suggest future research directions for developing more accurate models of users' behavior during EVA and building learning-aware VES. (Section \ref{sec: discussion})
\end{itemize}
\vspace{-2mm}
\section{Related Works}
% \begin{figure*}[!ht]
%     \includegraphics[width = \textwidth]{figures/subtask.png}
%     \centering
% \caption{Users' coming up with new subtasks as exploration progresses}
% \label{fig: subswitch}
% \end{figure*}
\noindent
\textbf{Static View of Exploratory Visual Analysis (EVA):} Researchers model user exploration behavior to improve how systems support EVA~\cite{gotz2009behavior, gotz2009characterizing,battle2016dynamic}. For instance, ForeCache uses Markov chain models to infer users' exploration goals from their interactions, which it uses to prefetch corresponding data regions~\cite{battle2016dynamic}. In contrast to our approach, these methods assume that users do not modify their strategies and follow a fixed exploration policy throughout the exploration process. 

\noindent
\textbf{User Learning in Data Querying:} Recent research using real-world keyword query workloads indicates that users learn to express specific and focused intents by modifying their keyword queries over time; furthermore, this learning behavior can be modeled using online learning algorithms~\cite{mccamish2018data}.
% We posit that the evolutionary nature of exploratory data analysis also shares strong parallels with online learning algorithms. 
% We also see researchers investigating user's learning behavior in the context of data querying. For example, \cite{mccamish2018data} models user learning in composing effective keyword queries in their long-term interaction with a data system. 
In \cite{cen2013reinforcement} researchers model users' evolving information-searching strategies from scholarly databases using Reinforcement Learning (RL). 
% They also show that users at different cognitive levels show different reinforcement patterns.
 Unlike in data querying, users often lack a predefined and concrete intent during EVA. Therefore, EVA presents a significantly larger action space requiring users to make more complex decisions.

% Unlike in data querying, users often lack a predefined and concrete intent during exploratory data analysis. Therefore, exploratory data analysis presents a significantly larger space of possible actions requiring users to make more complex decisions.

% \noindent
% \textbf{Expert-Aided Data Analysis.} Recent research uses RL trained on expert users' interactions to aid future analysts in similar data exploration tasks by automatically generating exploration sessions and relevant recommendations~\cite{seleznova2020guided, bar2020automatically}. In our work, we show users change their data analysis strategies over time to gain more rewards, i.e., information, even in a single exploration session. Hence, we propose modeling and adapting to users' learning strategies \textit{online}.

% \noindent
% \textbf{Low-Level Interaction-Based Models.} User models have been developed to evaluate the general usability of graphical user interfaces~\cite{jokinen2017modelling,ritter2000supporting, ivory2001state}. This learning process is often conceptualized initially as a trial-and-error approach, where users manipulate different interface components and observe the outputs to construct a mental model of the interface. With enough repetitions, users become experts in navigating the target interface~\cite{kasic1996toward}, be it digital (e.g., a search interface~\cite{teo2012cogtool}) or physical (e.g., a keyboard layout~\cite{jokinen2017modelling}). However, these models are too low-level to represent the abstract reasoning processes involved in exploring a complex dataset.

\noindent
\textbf{Understanding the User in Visual Data Analytics:} Models based on high-level user reasoning have been developed to aid users in visual data analysis. The closest examples of modeling user learning in visualization involve measuring the acquisition of knowledge in the form of insights~\cite{liu2014effects,battle2019characterizing,guo2015case,he2021characterizing}. However, insight-based analyses tend to focus on low-level metrics such as insight accuracy or insight generation rates~\cite{battle2019characterizing}, rather than testing whether the user is learning. Several works conceptualize a user’s reasoning process as they analyze a dataset~\cite{liu2010mental,patterson2014human}; however, these conceptualizations are unable to predict whether and how users learn from data analysis sessions. More recent work considers how users may update their prior beliefs in reaction to new data using Bayesian models~\cite{karduni2021bayesian,kim2019bayesian}. We consider a more general idea of testing whether users are learning as they encounter new data.

\section{Considerations for User Studies Selection}

\label{sec: considerations} 
The objectives and properties of exploration tasks may significantly influence how users engage with a VES ~\cite{battle2019characterizing}. 
Our selection of existing user studies by~\citeauthor{liu2014effects,battle2016dynamic, battle2019characterizing} for analyzing user behavior is by no means complete. 
But we believe they consider sufficiently many characteristics of exploration tasks to answer ``How Does User Behavior Evolve During Exploratory Visual Analysis?''
%learn from their past experiences.
\vspace{-3mm}
%---------------------
\subsection{Characteristics of Exploration Tasks}
\label{sec:charactetistics_of_exploration_task}
%---------------------
\underline{\textbf{Open-endedness:}}
Researchers categorize exploration tasks into three groups based on how clear users' objectives are~\cite{battle2019characterizing}.
%Researchers often categorize data exploration tasks based on how clear their objectives are~\cite{battle2019characterizing}.
%Based on which, one may place tasks on a spectrum ranging from those without a clear goal, to those with well-defined objectives.
%They usually recognize three types of exploration tasks in this spectrum. 

%---------------------
\noindent
{\bf Open-ended tasks} are
%are at one end of this spectrum, where the analyst initiates
exploration tasks without a clear intent or hypothesis. 
For example, in the {\it imMens user study} (Table \ref{tab:user_study_high_level}), participants search for \textit{interesting} information from large data. 
In these cases, information need is opaque, causing uncertainty in what and where to search. 
To successfully carry out this task, users might need to learn about the dataset during their exploration to be able to form relevant hypotheses and find interesting information.  
   
% For example, in some settings, users would like to identify any meaningful insights, salient patterns and outliers, correlations, trends, and other relevant observations in the dataset \cite{liu2014effects,reda2016modeling}.
%---------------------
\noindent
In {\bf goal-directed tasks}, analysts initiate exploration with a high-level goal or hypothesis. For example, in the {\it ForeCache user study} (Table \ref{tab:user_study_high_level}), participants' goal is to capture screenshots of 3 different U.S. regions with the highest snow coverage in a map visualization. Users may have multiple approaches to meet the goal, e.g., \textit{compare} snow coverage between various geographical regions, or \textit{search} an area in detail. So, they need to learn about the data to find the set of queries (or paths) that deliver the desired result (goal).

\noindent
In \textbf{focused tasks}, analysts have precisely defined goals and exploration paths. 
For instance, in the {\it Tableau user study focused task T2} (available in the appendix) 
% (shown in Table \ref{tab:tableau-table})
, users analyze how temperature changes over time. 
Users know what information to retrieve and which data area to explore as the question includes corresponding column names. 
Although, with the large data size, users may need additional interactions to find the provided locations. 
\\
%---------------------
\underline{\textbf{Prior Experience:}}
Users' prior knowledge about the dataset and familiarity with the exploration interface may influence the amount of information they should and will learn. 
In the imMens and ForeCache user studies, users have a limited time to get familiar with the dataset.
On the other hand, in the Tableau user study, the analysis tasks require users to explore an unseen dataset \cite{perer2008systematic}. 
%---------------------
\\
\underline{\bf Time Restriction:}
\label{sec: characteristics_time_restriction}
Users use the time between interactions to gather their thoughts, view data, draw comparisons, and plan their next steps \cite{battle2016dynamic,battle2019characterizing}.
As a result, time can be a vital factor in users' choice of learning schemes.
Our user study selection spans a diverse spectrum of time restrictions, ranging from \textit{no restriction} on Forecache to a \textit{30-minute limit} on imMens (Table \ref{tab:user_study_high_level}).
 \\
%---------------------
\underline{\bf Task Complexity:} 
\label{sec: cognitive}
The number of interface operations (action space) of the VES and the amount of information displayed has been shown to contribute towards task complexity and thereby impact users' exploration behavior in 
EVA \cite{gwizdka2010distribution, back2001model, lam2008framework}.
% information search \cite{gwizdka2010distribution, back2001model} and visual analytic tasks \cite{lam2008framework}. 
Among the systems used in our selected studies, ForeCache has the smallest action space, with only two options for navigating a 2D map, resembling Google Maps. imMens has four actions to query the imMens visualizations. 
In contrast, Tableau \cite{Tableau} is a more complex system with the largest action space for selecting, and filtering data.

\begin{table*}[h]
\resizebox{\textwidth}{!}{%
\begin{tabular}{|c|c|c|cllc|}
\hline
\multirow{2}{*}{\textbf{Characteristics}} &
  \multirow{2}{*}{\textbf{imMens user study \cite{liu2014effects}}} &
  \multirow{2}{*}{\textbf{ForeCache user study \cite{battle2016dynamic}}} &
  \multicolumn{4}{c|}{\textbf{Tableau user study \cite{battle2019characterizing}}} \\ \cline{4-7} 
 &
   &
   &
  \multicolumn{3}{c|}{\textbf{Task {[}T1, T2, T3{]}}} &
  \textbf{Task {[}T4{]}} \\ \hline
Exploration need &
  Open-ended &
  Goal-Directed &
  \multicolumn{3}{c|}{Focused} &
  Goal-Directed \\ \hline
Prior experience &
  15 minutes with datasets &
  Training subtask with dataset &
  \multicolumn{4}{c|}{5 minutes with a different dataset} \\ \hline
Time restriction &
  30 minutes per dataset &
  No &
  \multicolumn{3}{c|}{7 minutes per dataset} &
  5 minutes per dataset \\ \hline
Task Complexity &
  \begin{tabular}[c]{@{}c@{}}Interactive querying using \\ imMens actions on summarized plots\end{tabular} &
  \begin{tabular}[c]{@{}c@{}}Exploring 2D map\\ using pan and zoom\end{tabular} &
  \multicolumn{4}{c|}{\begin{tabular}[c]{@{}c@{}}Analyzing visualizations generated by \\ Tableau based on users' query on dataset\end{tabular}} \\ \hline
\end{tabular}%
}
\vspace{-2mm}
\caption{High-level dimensions in the exploration tasks of our selected user studies}
\vspace{-3mm}
\label{tab:user_study_high_level}
\small
\end{table*}
\vspace{-3mm}
\subsection{Methodology}
\label{sec:methodology}
We use the following methodology to analyze the evolution of users' exploration behavior across our selected studies. 
\vspace{-2mm}
\subsubsection{Overview of Exploration Task:}
We describe the analysis tasks in each user study and connect them with the characteristics introduced in Table \ref{tab:user_study_high_level}. 
Furthermore, we describe available user interface actions and associated interaction data utilized in our analysis.
% Furthermore, we briefly describe the user interface available to complete these tasks and the interaction data we use to analyze users' exploration behavior. 
% We describe each user study's tasks, goals, interface, and dataset(interaction log). We connect the task details with the characteristics  introduced in \autoref{tab:user_study_high_level}.
\vspace{-2mm}
\subsubsection{Formalizing the User Learning Problem:}
To understand users' approach to solving the task, we examine \textit{users' task activities} and identify the recurring strategies they exhibit. 
We define how users explore the dataset and make informed choices of exploration strategies, i.e., the \textbf{learning problem}. 
Then, we draw parallels between the proposed formalization and objective functions from well-studied online learning frameworks~\cite{auer2002nonstochastic,ontanon2013combinatorial,niv2009reinforcement}. As neither visualization researchers nor database researchers have tested for the exact objective function humans use during EVA scenarios, we explore various online learning frameworks tailored to the formalized learning problem.
\vspace{-2mm}
\subsubsection{Statistically Analyzing Behavior Evolution:} 
To better understand the evolution of user behavior, we perform statistical tests to check for changes in users' exploration strategies.
% \begin{example}
\label{example:nonstationary}
Let us revisit Example \ref{example:snowlevel}. 
To test her hypothesis, the analyst \textit{randomly} picks two different snowy mountain ranges and does lots of interactions to analyze them.
She successfully identifies some outliers for one of the regions but finds her approach time-consuming. 
Additionally, she learns that areas with multiple mountain ranges have higher chances of containing an outlier. \textit{Armed with this learning, she changes her approach} to enhance efficiency.
Initially, she zooms out to find candidate areas of interest and then decides to explore candidates in descending order of snow coverage.
This approach reduces the number of zoom operations and increases the chances of detecting outliers.
% \end{example}
% To accomplish this, we first formalize the exploration problems that users solve in our diverse selection of studies (\autoref{tab:user_study_table}).
% We use the formalized learning problem to generate users' exploration strategies.
By using statistical tests to analyze users' exploration patterns, we investigate changes in users' exploration behavior in different intervals.
% as they acquire more knowledge about the data. 
% Specifically, we divide the exploration session into smaller subsets and use statistical tests to determine if there are differences in users' exploration strategies between these mini-sessions. 
% Finally, we analyze these results and investigate how learning throughout exploration influences users to adopt fixed or changing strategies. 
% Finally, we analyze these results and investigate how learning throughout exploration influenced users to adopt a fixed (i.e., stationary) or changing (i.e., non-stationary) strategy. 

% , facilitating the analysis of the affects of user learning.

% \textbf{Example:} Using a flight performance dataset \cite{FAA1}, an analyst compares the number of delayed flights between Alaska and Delta. 
% This task requires minimal exploration, where she can focus on a single imMens visualization (\autoref{fig: immens_interface}) and analyze it to answer. 
% Such tasks may not be enough to give a general understanding of the data, which can change users' exploration strategies. 

% \noindent

% \noindent
\vspace{-2mm}
\subsubsection{Evaluated Learning Algorithms:} 
% We utilize scholarly works in neuroscience, cognitive psychology, economics, game theory, etc., \cite{niv2009reinforcement, niv2012TD, roth1995learning, glimcher2011dopamine} to find algorithms that can replicate users' learning behavior. 
In this paper, we address the absence of a comprehensive investigation of human learning during EVA. To fill this void, we employ human learning algorithms commonly utilized in economics \cite{bush1953stochastic, roth1995learning}, game theory \cite{tamura2015win}, artificial intelligence \cite{mccamish2018data, cen2013reinforcement, zhang2013forgetful}, and cognitive psychology \cite{niv2009reinforcement, niv2012TD, glimcher2011dopamine}. 
We briefly describe the algorithms and justify their selection for modeling users' exploration behavior in our selected studies. 
We establish a consistent measure of \textit{how well these learning algorithms can adapt to users' exploration strategy} by \textbf{empirically evaluating their performance in predicting users' future actions}.
\section{ForeCache User Study}
\label{sec: forecache}

\subsection{Overview of Exploration Task}
\label{sec: forecache_overview}
\noindent
\textbf{Analysis Task:} 
In the ForeCache user study \textbf{task}, participants 
% (students, post-doctoral researchers, and faculties from UW and UCSB) 
explore a map visualization containing information on snow coverage in the US (Figure \ref{fig: forecache-interface1}). The task \textbf{goal} is to look for areas with high snow coverage on the map. Users are expected to provide screenshots of three such areas as \textbf{results}~\cite{battle2016dynamic}.

\noindent
\textbf{Characteristics: }
We show in Section~\ref{sec:charactetistics_of_exploration_task} how analysts know the data features to search for but have unclear path ideas, making this EVA task goal-directed. 
Given the limited expected results (3 screenshots), there is no \textit{time restriction}. Users complete a small training subtask before the task, so they had limited \textit{prior experience} with the visualization tool.
% The task goal precisely defines which data features(snow coverage) to search for and what data areas (with high snow coverage) to find. Users may have multiple paths to meet the goal, e.g., compare
% snow coverage between various geographical regions or search an
% area in detail. Therefore, with clear goals but an unclear path, the exploration task is \textit{goal-directed}. Given the limited expected results (3 screenshots), there is no \textit{time restriction}. Users complete a training subtask before the task, so they had some \textit{prior experience} with the visualization tool.

\noindent
\textbf{Interface and Interaction Log:}
ForeCache \textbf{interface} displays snow coverage (\textbf{snow level}) in shades of blue based on Normalized Snow Index (\textit{NDSI}) calculations from NASA MODIS dataset~\cite{RITTGER2013367}. Snow levels are in the range [0: low snow level, 1: high snow level]. 
The interface supports \textit{pan} to explore up, down, left, right, and \textit{zoom} to observe snow coverage in different levels of detail (\textbf{zoom levels}). There are six different zoom levels, from 0 (coarse) to 6 (fine-grained). 
The interface's actions compel users to incrementally and \textit{sequentially} retrieve only a fraction of the underlying dataset. 
For example, the user cannot directly jump from zoom level 0 to 4; she has to incrementally go through levels 1, 2, and 3. %, similar in functionality to Google Maps. 
The \textbf{interaction log} from the user study, NDSI 2D Interaction Dataset~\cite{battle2017behavior}, comprises data exploration logs of 20 participants.

%---------------------------------------------

\begin{figure*}[!htbp]
    \includegraphics[width = \textwidth]{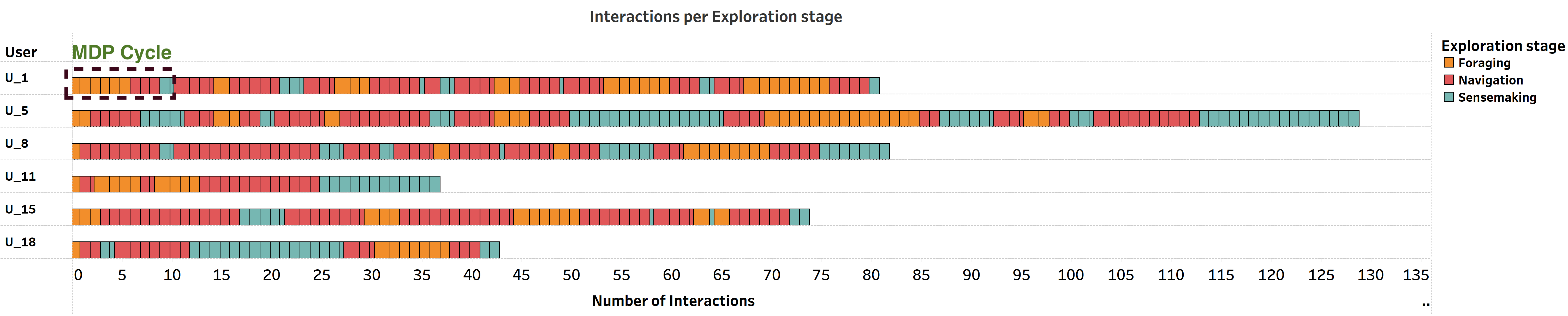}
    \centering
\vspace{-7mm}
\caption{6 random users' changing preference for three exploration stages: \textit{Foraging}, \textit{Navigation}, and \textit{Sensemaking}.}

\captionsetup{justification=centering}
\vspace{-3mm}
\label{fig: forecache-states-switch}
\end{figure*}

\subsection{Formalizing User Learning Problem}
\label{sec:forecache_modeling}
\textbf{User Activities and Learning Problem:}
\label{sec: forecache-user-activities}
To comprehend how users are evolving in their exploration behavior, we need to capture users' strategies for solving the task. The raw ForeCache action space, e.g., zooming and panning, may not provide a meaningful categorization of users' exploration strategies \cite{battle2016dynamic}.

% Researchers in visual data exploration address this challenge by 
Visualization researchers address this challenge by 
characterizing users' strategies using high-level exploration stages \cite{gotz2009behavior, saha2019leveraging}. Therefore, we follow \cite{battle2016dynamic} and use the three stages of exploration based on information foraging theory \cite{pirolli2005sensemaking}.

\noindent
\textbf{\textit{User Actions:}} To explore the dataset, the user starts \emph{foraging}, i.e., looking for interesting patterns in the data at coarse zoom levels Figure (\ref{fig: forecache-interface1}). Her goal is to \textit{generate hypotheses} for potential screenshot candidates in high snow-covered areas. 
Then, she gradually moves to fine-grained zoom levels, i.e., {\it navigates} to more detailed views of snow coverage. 
Having identified an area of interest at a fine-grained zoom level, she starts \emph{sensemaking}, i.e., comparing snow coverage in detail. To test her hypothesis she verifies if this candidate area has high snow coverage. 
However, the user is not limited to analyzing a single area; she may also \textit{navigate} back to a bird-eye view to \emph{forage} for new snow areas or hypotheses.

\noindent
\textbf{\textit{Exploration Stages:}} In Table \ref{tab:mapping_actions_goals}, we present the three exploration stages derived from users' activities, i.e., low-level actions and goals: Foraging, Navigation, and Sensemaking. As shown in Figure \ref{fig: forecache-states-switch}, the sequential dependence between the stages of exploration (Foraging-\emph{then}-Navigation-\emph{then}-Sensemaking) makes users' exploration process suitable to be formalized as a Markov Decision Process (MDP).

% At each interaction, users should decide that given their current stage (Foraging/ Navigation / Sensemaking), whether they should stay at the same stage or move to another one to maximize the chances of discovering high snow regions and target zoom-level.

%---------------------------------------------
% \subsection{Modeling User Learning}

%discounted future reward. The discount factor accounts for future rewards being worth less than immediate rewards.

% Unlike in bandit problems, MDPs are designed for problems where observations are based on specific features of the environment. And there exist actions that help an agent move between these states.

\begin{table}
  \small
  \caption{Mapping actions and goals to exploration stages}
  \label{tab:mapping_actions_goals}
  \begin{tabular}{p{1.8cm} p{2.4cm} p{3.0cm}}
    \toprule
    \textbf{Stage} & \textbf{Goal} & \textbf{Low-level action} \\
    \midrule
    Foraging & Generate hypotheses & Pans at coarse Zoom Level \\
    \hline
    Navigation & Navigate & Zoom in/out \\
    \hline
    Sensemaking & Test hypotheses &Pans at fine-grained Zoom Level \\
 
    \bottomrule
  \end{tabular}
  \vspace{-4mm}
\end{table}

%------------------------------------------------------
\noindent
\textbf{Modeling Exploration Using MDP:\nopunct}
\label{decision-problem-mdp}
We formalize users' exploration approach as MDP's state, action, and reward.
Background for MDP is available in the Appendix \ref{subsection: mdp}.

\noindent
\textbf{\textit{States}} We extend the stages in Table \ref{tab:mapping_actions_goals} as MDP states.

\begin{figure}[!htbp]
\centering
\vspace{-3mm}
\includegraphics[scale=0.27]{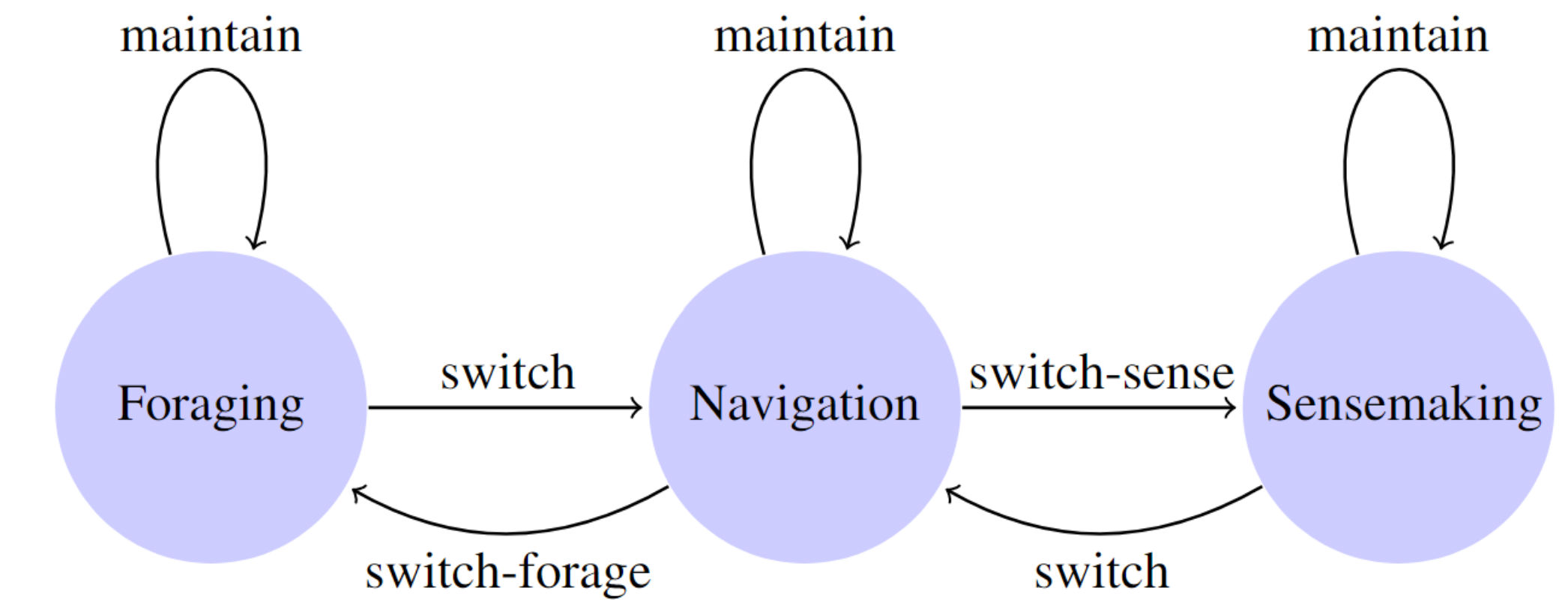}
\centering
\vspace{-3mm}
\caption{All possible transitions between the 3 MDP states}
\vspace{-5mm}
% \Description{Displays a MDP graph. With nodes representing the 3 states -Foraging, Navigation, Sensemaking, and edges representing the possible transitions. Cyclic edge from a state to itself is annotated as `maintain’ action. Edge from the Foraging node to the Navigation node is `switch,’ same for Sensemaking to Navigation. Edge from Navigation to Foraging is `switch-forage,’ and Navigation to Sensemaking is `switch-sense’. }
\label{fig:mdp}

\end{figure}

\noindent
\textbf{\textit{Actions:}} In Sensemaking and Foraging states, a user can perform two actions: \textit{switch} or \textit{maintain} (Figure \ref{fig:mdp}). For instance, by picking action \textit{switch}, the user changes her exploration strategy from Sensemaking to Navigation (and vice versa). In the Navigation state, the user's \textit{switch} action can be further divided into \textit{switch-forage}, taking them to the Foraging state and \textit{switch-sense} for transitioning to Sensemaking. When the action is \textit{maintain}, the user maintains the current exploration strategy and stays in the same state. 

\noindent
\textbf{\textit{Reward:}} On instigating these transitions, user receives feedback from the interface environment. This feedback is a measure of how successful the user's actions are in achieving the task requirements of taking screenshots at high zoom levels and looking for high snow coverage. We extend this response from the interface as the reward in the MDP environment; \textit{Reward = Snow level $\times$ Zoom level}

\noindent
\textbf{\textit{Learning Problem:}} At each interaction (\(t \in [1, T]\)), users should make a learned choice, given their current state, whether they should stay there or move to another one to maximize the chances of discovering high snow coverage. Similar to an MDP agent, a user's objective during exploration is to learn an optimal policy (\(\pi : \textnormal{State} \rightarrow \textnormal{Action}\)) that maximizes the expected sum of future reward,
\(\max_{\pi} \boldsymbol{E} \left[ \sum_{\textnormal{interaction}=t}^{T} \textnormal{reward}(\pi, \textnormal{interaction}) \right]\).

\noindent
\textbf{Learning in an MDP Setting:}
\label{forecache_userlearning}
One approach to quantifying behavioral shifts is to analyze the improvement in accumulated rewards. For some users, we observe an increasing trend in \textit{rewards over interactions}, indicative of users finding higher snow-covered regions as they progress through the task. 
Upon analyzing user behavior at different stages of the task, we observe a notable difference between early and later exploration interactions. Similar to most RL agents~\cite{sutton1998}, users tend to explore more than exploit at the start. 
It involves staying in the Navigation state: switching between different zoom levels and exploring different snow areas. 
After learning more about the dataset, users can exploit each candidate area for detailed analysis.
%In addition, we also note that idle time between interactions is highest at the beginning, possibly because users are learning the visualization overview outside of their logged-interactions. This also highlights the \textit{time between interactions}(\autoref{sec: characteristics_time_restriction}) users may leverage in a task without a set time limit.  %cite Waldo ? where these metadata are used to categorize users%

 % \leilani{I like this!}

% \begin{figure}[!h]
% \centering
% \includegraphics[scale=0.8]{figures/ROI-summary-short.png}
%   \caption{Increasing Reward and Decreasing Idle Time over Exploration for three random users with low reward and high idle time in the first ROI}
%   \label{fig:forecache-reward-update}
% \end{figure}

\subsection{Statistically Analyzing Behavior Evolution}
\label{forecache-ns}
% Based on our methodology outlined in subsection 3.2, we test for changes in users’ exploration behavior. 
We split each participant's exploration session into two partitions $S_1$ and $S_2$.
We chose to use a 50-50 split, as it represents the most fundamental level at which we would expect to observe changes in strategy. We extract the probability of a user preferring a specific state over other states. Given that Navigation accounts for over $50\%$ of user interactions and state preferences are dependent, we calculate the probability of users preferring Navigation between $S_1$ and $S_2$. Then we use the Wicoxon-Signed-Rank test~\cite{woolson2007wilcoxon} to test if all users' preference changes. 

We find a significant difference in all users' state probabilities of preferring the Navigation state over other states (Statistic:28 p-value:0.002 ). This result provides quantitative evidence of users' changing preference for exploration stages, as visualized in Figure \ref{fig: forecache-states-switch}.

%--------------------------------------------------------------------

\subsection{Evaluated Learning Algorithms}
\label{forecache_algorithms}
Our choice of algorithms includes human learning algorithms from neuroscience and cognitive psychology. We also include some heuristic algorithms, Win-Stay-Loose-Shift(WSLS), Greedy and Random (Section \ref{appendix: forecache_algorithms}). The \textbf{objective} of all chosen algorithms is to maximize reward. We avoid using overly sophisticated algorithms that require a lot of information and computation as research in \cite{vandekerckhove2015model} shows simple ones generally model human behavior more accurately. 

We use value-based reinforcement learning (RL) algorithms, Q-learning (QLearn) and State-Action-Reward-State-Action (SARSA) as such algorithms have been used in explaining human learning \cite{glimcher2011dopamine}, and decision-making  behavior~\cite{niv2009reinforcement,niv2012TD,daw2011modelbased}.
Alternatively, some cognitive psychology researchers argue for using \textit{value-free or policy-gradient} RL methods, or a combination of value-based and policy-gradients in explaining human behavior~\cite{bennett2022model} and learning in brain~\cite{joel2002actorcritic}. Reinforce and ActorCritic are used to represent such methods. Additional information about these algorithms and the justification of their selection is available in the appendix.

% \begin{equation}
% V(s_t) \leftarrow V(s_t) + \alpha \left(r_t + \gamma V(s_{t+1}) - V(s_t)\right)
% \label{eqn: actorcritic}
% \end{equation}
% Where $V(s_t)$ is the estimated value of state $s_t$. This model has $\gamma$ and $\alpha$ as hyperparameters.The neural network's final output are passed through a \textit{softmax} function to get a probability distribution over possible actions.

% \subsubsection{Experiment procedure: }
% We begin with simple heuristics, Naive, Win-Stay-Lose-Shift (WSLS), and Greedy algorithms(\autoref{sec: tableau_methods}), followed by value-based algorithms QLearning and SARSA. We then examine a policy-based algorithm Reinforce and explore the potential benefits of combining value and policy learning using the Actor-Critic algorithm.For all algorithms, we follow the same procedures for model fitting and hyperparameter tuning as described in \autoref{sec:tableu_empirical_evaluation}
\vspace{-3mm}
\subsection{Performance Evaluation}
\label{sec: forecache_results}
\noindent
\textbf{Evaluation Procedure:} 
\label{sec: evaluation_procedure_foreCache}
User learning may vary at different stages of EVA. To capture this nuance, we train our algorithms using a varying range of training data, from a lower limit (users finish learning early on) of $5\%$ to an upper limit (users keep learning till the end) of $90\%$ of all user interaction data. 
Using the training data within each threshold, we find the optimal hyperparameters and use them to train our models.
Using remaining data, we evaluate each algorithm's performance in predicting \textbf{What 
 \textit{action}(switch or maintain) a user will use in her next interaction.}.
%first state your results/finding/argument, then your conclusion, ...
%RESULT THEN ARGUMENT

\begin{figure}[!htbp]
\centering
\vspace{-3mm}
\includegraphics[scale=0.15]{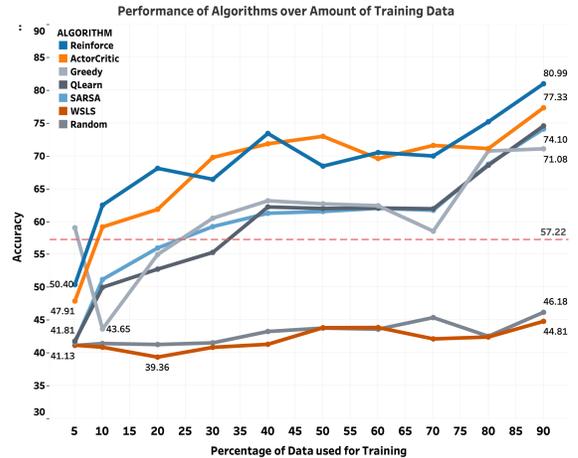}
\vspace{-3mm}
\caption{Algorithm performance on different thresholds}
\vspace{-3mm}
\label{fig:performance-data-level}
\end{figure}

\noindent
\textbf{Evaluation:}
\label{sec: forecache_results_user_data}
In Figure~\ref{fig:performance-data-level}, algorithms' accuracy increases with more training data. It is similar to how participants learn from new experiences during the task (Section \ref{forecache_userlearning}).

% \subsubsection{Top Algorithm per User:}
% We also evaluate the performance of all algorithms based on how well they predict the next action for each of the 20 users. 
Among the 7 algorithms, Reinforce has the highest prediction accuracy for $70\%$ of the users, followed by Actor-Critic ($20\%$) and Greedy ($10\%$). It suggests that a single algorithm may capture most users' learning, despite users' having differences in exploration strategies (Figure \ref{fig: forecache-states-switch}).

% This suggests although users have different degrees of non-stationarity (\autoref{forecache-ns}), a single algorithm may capture most users' learning.
 % Our findings on (a) differences in nonstationarity between users from \autoref{forecache-ns}, (b) unrestricted and varying time spent on the task from \autoref{tab:user_study_high_level}, and even (c) differences in the number of interactions and ROI cycles from \autoref{fig: forecache-states-switch},

%At the user level, hyperparameters such as epsilon and learning rate also play a crucial role in determining the performance of each algorithm. However, users may change their behavior over time, so fitting the model on one set of hyperparameters may not yield the same performance in testing.

%At the data level, the performance of each algorithm may be affected by the amount of available data, the quality of the data, and the variation in user behavior.
% In an online setting, a good model should be able to learn quickly and accurately with minimal data. Therefore, we evaluate the performance of our algorithms after learning on different amounts of user interaction data.  

%shows the best-performing model at each threshold, i.e., the model that best matches each user's actions/behaviors.

%For example, a user may have a preference for certain types of items or stores, and the Actor-Critic model can learn and exploit these preferences to improve performance. The figure shows an example of one contextual preference and the performance of one user.

%----------------------------- RESULT -----------------------------
% \begin{figure}[!htbp]
% \centering
% \includegraphics[scale=1.4]{figures/all-algos-global-performace.png}
% \caption{Performance of discussed MDP based algorithms across all users and thresholds}
% \label{fig: globalAccuracy}
% \end{figure}
% \vspace{-3mm}
\noindent
\textbf{Overall Results:} 
When discussing algorithm performance, we report the aggregate results 
from Figure \ref{fig:performance-data-level} with parenthesis alongside their name. 
Simple heuristics [Random ($43\%$) and WSLS ($42.1\%$)] were not enough to capture the nuances of user learning, failing to keep up with users' evolving exploration behavior.
Policy gradient-based algorithms (Reinforce ($68.6\%$), Actor-Critic ($67.3\%$)) are better suited for capturing action choices because they provide a probability distribution over all actions without requiring fine-tuned exploration-exploitation parameter ($\epsilon$). 
They can also approach deterministic policies asymptotically; as~\citeauthor{sutton1998} note, it is challenging to achieve with $\epsilon$-greedy and action-value methods. 
The low task complexity with limited sequential actions and well-defined goals of the goal-directed task (Section \ref{sec: forecache_overview}) also supports policy gradient as a more suitable learning scheme. Learning a policy with a value-based approach (QLearn ($59.1\%$), SARSA ($59.7\%$)) is a more complex two-step process where users first need to form value functions based on observed snow coverage and only then learn specific probabilities for taking actions (policy)~\cite{bennett2021value}.
Lastly, we observe that the simpler Reinforce algorithm performs similarly, if not better, than the Actor-Critic, which justifies our inclination toward simpler algorithms to understand user learning.

\vspace{-3mm}
\section{imMens user study}
\label{section: datasets}
\subsection{Overview of Exploration Task}
\noindent
\textbf{Analysis Task:}
\label{section:dataset-imens}
In this user study by \citeauthor{liu2014effects} 16 participants 
% (professionals in the Bay area experienced in data exploration) 
perform open-ended tasks on two datasets.
They report any \textbf{interesting findings}, which the authors define as surprising events, data abnormalities, confirmations of common knowledge, or intuition; in other words, \textbf{\textit{new data or statistics}} that the user did not know or was unsure of beforehand. 
The users explore following datasets: (a) travelers' check-in data on \textit{Brightkite}, a location-based check-in service, and (b) US domestic flight performance data.

\noindent
\textbf{Characteristics:}
The task description and a few {\it `interesting findings'} examples give participants a vague idea of what they need to find.
However, they lack precise knowledge of what to search for (i.e., data abnormalities, surprising events, etc.).
Moreover, they are uncertain about the availability of new information and search locations.
Therefore, without a clear goal of what to find and exactly which part of the data to explore, they proceed to complete this \textit{open-ended} task.

The participants have a 30-minute \textit{time restriction} per dataset but can quit analysis if they feel there is nothing more to discover.
Additionally, 15 minutes of \textit{prior experience} with the datasets and interface may impact their learning to a certain degree.

\begin{figure}[!htbp]
    \includegraphics[scale=.14]{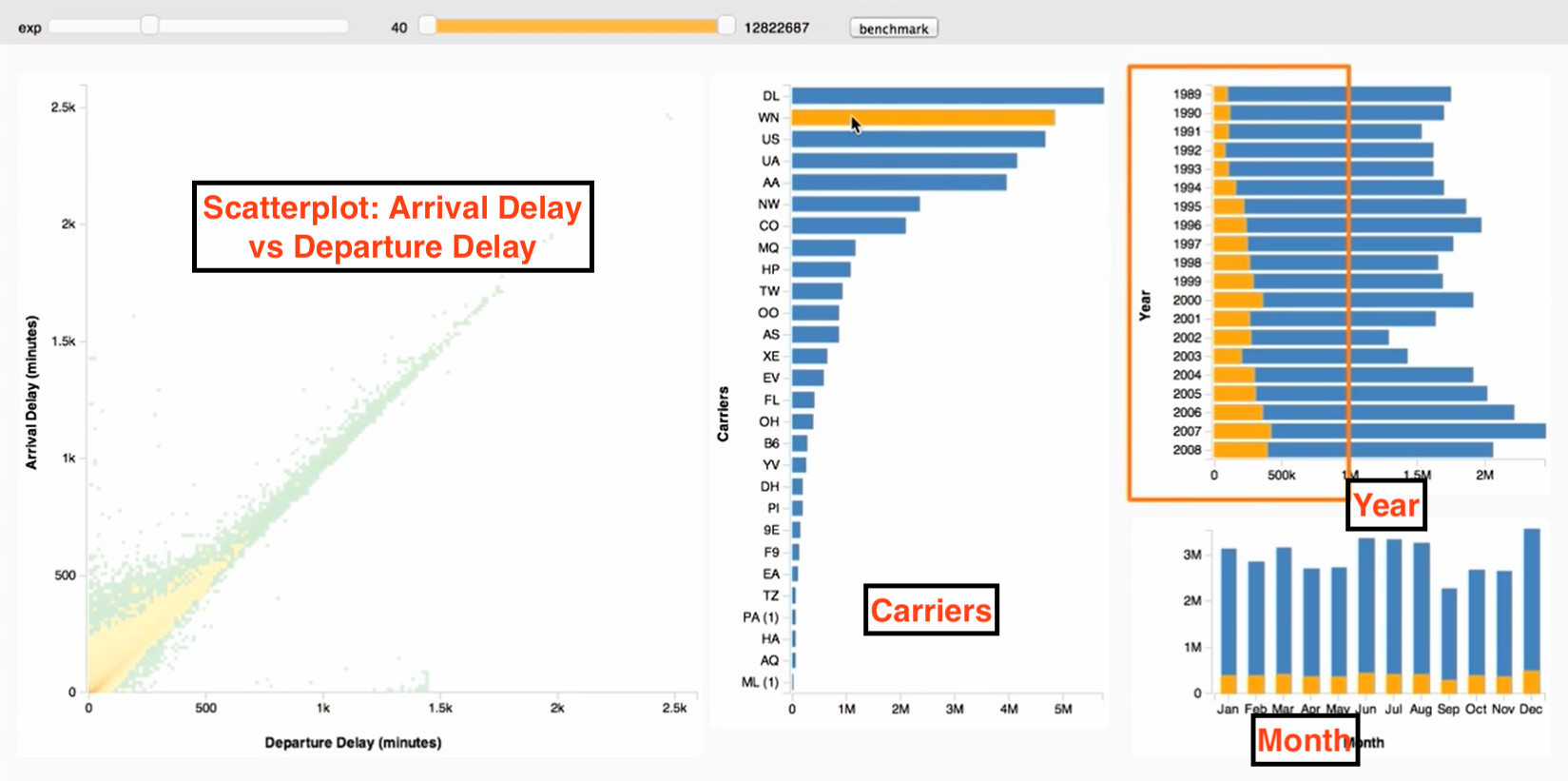}
    \centering
\caption{imMens user exploring flight performance data}
\captionsetup{justification=centering}
\label{fig: immens_interface}
\vspace{-3mm}
\end{figure} 

\noindent
\textbf{imMens Interface:}  
\label{subsubsec: imMens_interface}
It displays four visualizations for the flight performance dataset (Figure \ref{fig: immens_interface}). 
Participants can explore them using \textit{brush \& link, pan, zoom, and select}. 
The visualizations are (1) Scatter plot: 2D plot showing the relationship between arrival and departure delays. (2) Carriers: bar chart with flight information of various US airline carriers. (3) Year: bar chart, and (4) Month: histogram with flight information for the respective time units.

\begin{comment}
These visualizations are named \textit{linked visualizations} because operating on one cascades changes to other visualizations accordingly.
For example, a user selects Year = 2003 in the visualization Year in Figure \ref{fig: immens_interface}.
Now, she will see the number of flights in 2003 for different US airlines, the distribution of those flights across months, and an associated scatter plot in corresponding linked visualizations.
\end{comment}
Similarly, upon uploading the travelers' check-in dataset, imMens presents five visualizations in its interface. They are (1) a multi-scale geographical heatmap depicting travelers' check-in locations worldwide. (2, 3, 4) Three histograms aggregating the number of check-ins by day, month, and hour, and (5) a bar chart showing the top 30 travelers with the highest check-ins in the current geographic bounding box.
\\
\noindent
{\bf imMens Interaction Log:}
contains users' raw interactions (e.g., zoom, pan, brush, etc.) with imMens.
Additionally, it has users' verbal feedback, where they explain their actions, findings, and reasonings, e.g., what type of information they want to find, if they have discovered anything new, etc.

\begin{figure*}[!t]
    \includegraphics[scale=.5]
    {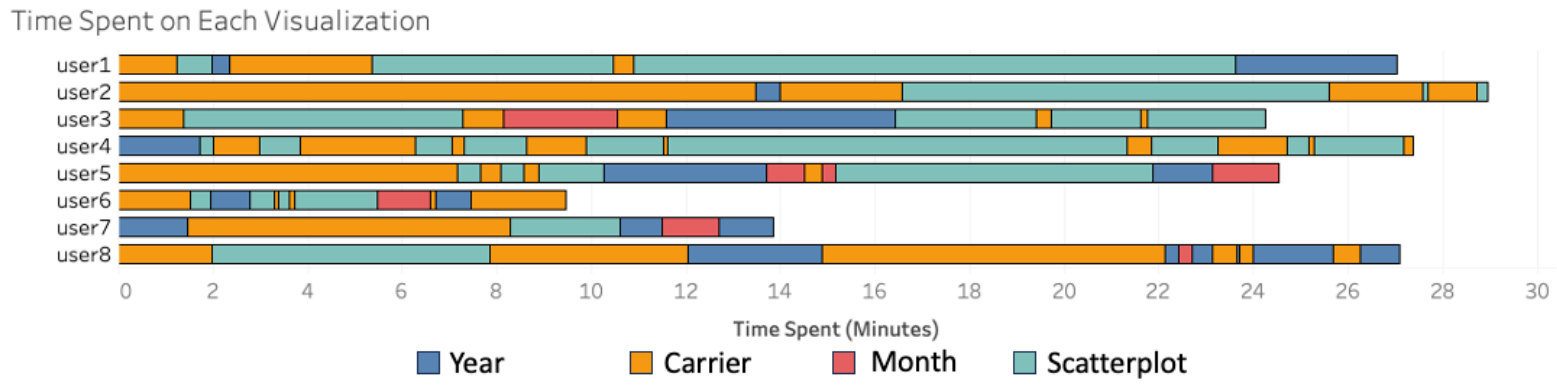}
    \centering
    \vspace{-3mm}
\caption{Users focus on different visualizations in the imMens interface while exploring the flight performance dataset.}
\captionsetup{justification=centering}
 \vspace{-6mm}
\label{fig: time_spent_tableau}
\end{figure*} 
   
%\label{sec: imMens_decision_problem}

% In this user study, \citeauthor{liu2014effects} wants to explore the effects of additional delay in EVA.
% So, the participants explored one of the datasets using the imMens interface with an added 500 ms response latency. 
% As we focus on analyzing users' exploration behavior in EVA, we use the interactions without additional latency.

% \input{5_1_iMens_modeling_decision_problem.tex}
\subsection{Formalizing User Learning Problem}
\noindent
\textbf{User Activities:}
\label{sec: users_decision_problem}
In open-ended exploration, the user aims to learn new information without any specific exploration path or goal to complete a task.
Users unravel new information by interacting with the imMens interface.
After each interaction, the user analyzes the updated visualizations for new information.
%When a user operates on a visualization, she learns additional information from other linked visualizations. 
%So, after interacting with the visualization, the user analyzes the information on the interface.
If the user discovers any findings, she reports them before continuing her exploration.
In each step, \textit{she makes a learned choice about which visualization to interact with to discover new information from that data area.}

In Figure \ref{fig: time_spent_tableau}, we show the chronological order of users' interactions with the linked visualizations while exploring the flight performance dataset.
We can see that users perform multiple consecutive interactions with a visualization to find new information from a specific data area before switching.
% We can see that users perform multiple consecutive interactions with a visualization to find new information, e.g., information about specific data areas, before switching to a different visualization/data area.

\noindent
\textit{\textbf{Actions:}}  
In each step $t$, the user selects one visualization $v_t$ to interact with from K visualizations in the interface.

\noindent
\textit{\textbf{Reward:}} 
The information obtained from the interface serves as a reward for interacting with the visualization.
The reward value depends on the effectiveness of users' decision to interact with the selected visualization towards new information.
In our experiments, we utilize the feedback log to evaluate the effectiveness of each user's learned decision of visualization selection and assign a reward, $R \in [0, 1]$.
When the user uncovers a piece of new information by interacting with a visualization, they receive the highest reward ($R = 1$).
The feedback log allows us to identify meaningful information discoveries because users report their findings as part of the study protocol, such as answers to the question she was searching for, generalization to the observed patterns, and confirming a hypothesis.

\textit{Generating hypotheses or questions} reveals a potential shift in user intent to uncover new information or patterns.
We assign a reward ($R = 1$) for these cases.
Besides these rewarding interactions, users learn about the data by \textit{observing} the visualizations.
Although users may not report any interesting findings during these interactions, they are still crucial for generating intent and progress toward the desired result.
Therefore, we assign a small reward ($R = 0.1$) to such instances.
\citeauthor{liu2014effects}'s categorization of user feedback 
% into seven groups, including observation, generalization, hypothesis, question, etc., drawing from prior EVA research.It 
helped us ensure a consistent reward assignment for the visualization selections.  
We assign $R=0$ to instances where the users' visualization selection does not affect their goals, as they were configuring the interface.
% \vspace{-3mm}

\noindent
\textbf{Learning Problem:}
\emph {In each step $t$, a user makes a learned decision: (a) continue focusing on a specific visualization, or (b) pick a different one}. Here $t\in[1, ..., T]$, where 1 is the starting point, and $T$ represents the final time step.
This decision may change based on their gained knowledge from the data and information need, which may influence the picked action $v_t$. 
The user learns the optimal policy by optimizing the objective function that maximizes the expected new information,
% expected interesting findings,  
$\sum_{t}^{T}\mathbb{E}[\mathbb{R}eward]$. 
A user following the optimal policy $\pi^*$ may continue action $v_t$ when generating a hypothesis or seeking answers.
Or if she believes $\pi^*$ will lead to more opportunities for intriguing discoveries. 
Conversely, after reporting an insight, the user may decide whether to switch to a different visualization. 
The user's perception of thoroughly exploring the current visualization, its potential for discovering new information, and the presence of a new hypothesis to test may influence such a switch.

\vspace{-3mm}
\subsection{Statistically Analyzing Behavior Evolution}
\label{sec: iMens_ns}
We statistically evaluate users' preference changes in interacting with specific visualizations between two exploration stages. Each participant's exploration session is divided into equal halves and the probability of selecting particular visualizations in each partition is analyzed. Complete results of the tests are available in Appendix \ref{subsec: imMens_wilcox}. Let's focus on the travelers' check-in dataset, where users majorly interacted with geographical heatmap visualization (geo-plot). 
We use the Wilcoxon-Signed-Rank test and find no significant change in user preference for interacting with geo-plots between the partitioned exploration phases (statistic=6.0, p-value=0.11). 
It aligns with users' reported insights being predominantly related to the geographical areas travelers are from, e.g., the continents, US states, and countries, and required users frequent interaction with the geo-plot.
For none of the visualizations, we see any significant difference in users' behavior significantly evolving during this open-ended exploration.  
\vspace{-3mm}
\subsection{Evaluated Learning Algorithms}
\label{sec: algorithms-openended}
We repeat the same baseline, heuristics, and $\epsilon$-Greedy based algorithms (Appendix \ref{sec: tableau_methods}) to model users’ evolving
exploration behavior for this user study. 
In addition, we use mortal-arm bandit and contextual bandit algorithms, commonly utilized in modeling user behavior in online advertising, as detailed in Appendix \ref{appendix: imMens_algorithm}. Here, each visualization is considered an arm in the multi-arm bandit setting.

\noindent
{\bf Evaluation Procedure:} 
% To account for potential variations in user learning during different stages of data exploration, 
Similar to Section \ref{sec: evaluation_procedure_foreCache},
we evaluate the algorithms using varying amounts of training data. 
 and prediction accuracy on users' next choice of visualization.
% \vspace{-8mm}

% \input{5_5_iMens_results_discussions.tex}
\begin{figure}[!htbp]
\centering
% \vspace{-3mm}
\includegraphics[scale=.25]{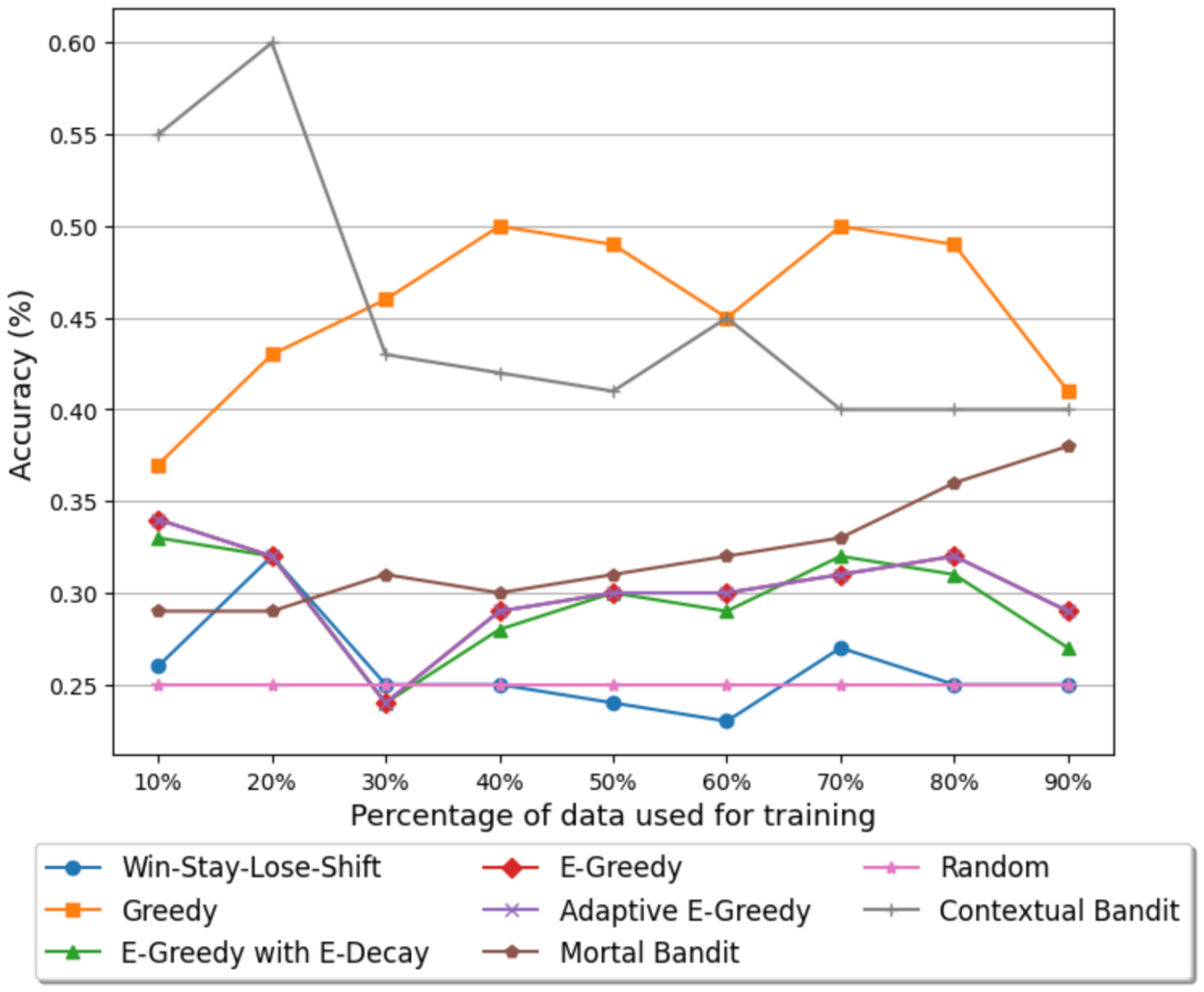}
\vspace{-3mm}
\caption{Algorithm accuracy on different training threshold for flight performance dataset}
\vspace{-7mm}
\label{fig: accuracy1_faa}
\end{figure}
\vspace{-3mm}
\subsection{Results}
\label{sec: imMens_results}
\begin{figure}[!htbp]
\centering
\includegraphics[scale=.25]{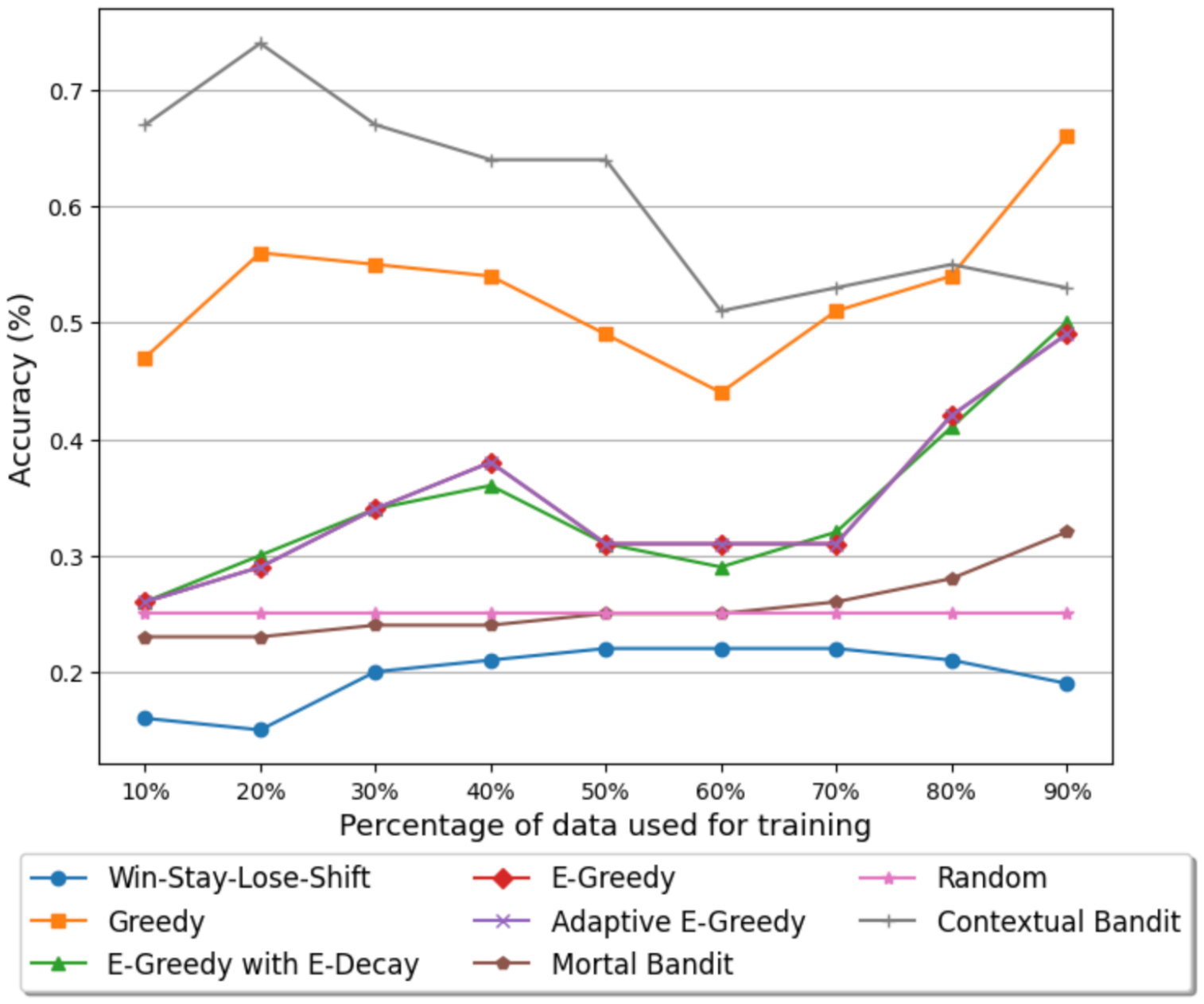}
% \caption{Accuracy of algorithms after being trained on different percentages of data}
\vspace{-3mm}
\caption{Algorithm accuracy on different training threshold for travelers' check-in EVA}
% \vspace{-4mm}
\label{fig: accuracy1_brightkite}
\end{figure}

Figure \ref{fig: accuracy1_faa} and \ref{fig: accuracy1_brightkite} show that the greedy algorithm performs much better than $\epsilon$-greedy based approaches with $49\%$ accuracy.   
Greedy works well as user's visualization preferences do not significantly differ and they often continue interacting with the same visualization to discover multiple new information and generate hypotheses related to a particular data area (Figure \ref{fig: time_spent_tableau}).
Contextual-bandit (CB) performs the best with an average of $53\%$ accuracy. In each step, CB observes a user's raw action to predict the user's next choice of visualization. However, doing it did not create any significant improvement over the greedy algorithms, as CB did not find a significant correlation between the raw action space and the picked visualization. 

On the other hand, the $\epsilon$-Greedy struggles to adapt quickly to changes in users' visualization choices as they navigate to different data areas. 
From Figure \ref{fig: time_spent_tableau}, we see users' interaction time with a particular visualization varies throughout the exploration session, which makes it hard to determine appropriate hyperparameters. 
As the user learns, their rate of exploration changes during EVA. $\epsilon$-Greedy algorithm's hyperparameter ($\epsilon$) trained on a specific EVA segment doesn't guarantee the user will have the same exploration rate for the remaining session. Moreover, fine-tuning the decay parameter in real-time is challenging, especially in non-stationary environments, without yielding substantial benefits.
Adaptive $\epsilon$-greedy and mortal bandit algorithms face the same limitation due to their reliance on hyper-parameters, step size, and lifetime, resulting in poor performance. 

Simple heuristics and algorithms like WSLS, Greedy, and $\epsilon$-Greedy exhibit limited adaptability to user behavior in intricate data exploration. Conversely, complex approaches such as adaptive $\epsilon$-Greedy and mortal bandit lack the flexibility to adjust to users' future visualization preferences.

Certain users prefer to explore different visualizations within shorter timeframes during the later stages. Such behavior leads to performance declines in those periods, as the learned strategies from the initial stage's training data do not encompass such patterns. 
This trend is evident in Figure \ref{fig: accuracy1_faa}. However, greedily choosing recently rewarded visualization leads to relatively improved performance during those intervals.
Meanwhile, similar to the ForeCache user study, some users start by exploring different data areas to evaluate the available information and subsequently leverage it for their benefit.
As a result, the hyperparameters trained in the initial exploration phase cannot cope with the rate of changes in visualization selection. 
% Interestingly, we observe a decrease in algorithm accuracy beyond the 60\% training threshold, which contrasts with the conventional trend of accuracy increasing with larger training datasets.
% This is due to users opting for different exploration policies at the later stages of their session compared to the initial. 
% , we note that users tend to explore more extensively in the later stages of the initial exploration phase.
% This behavior may be attributed to users initially concentrating on a single visualization to extract as much information as possible within the given time limit.
% However, upon completing the initial exploration, if users perceive available time, they may go back and forth on different visualizations for more patterns.

% offline 
% algorithms have different appoaches to objective functions so there is no unified way, users are trying to do this ...
% users are regret ..

%faa [0.55 0.6  0.43 0.42 0.41 0.45 0.4  0.4  0.4 ]
% [0.66618272 0.74329935 0.67195423 0.64373908 0.63568929 0.51021764
%  0.52888804 0.5491849  0.52868268]
%brightkite
\vspace{-6mm}
\section{Discussion}
\label{sec: discussion}
To enrich the HCI and visualization communities' current understanding of the evolution of user exploration and its modeling process during EVA, we identify the common themes across popular studies, analyze users' behavior, and empirically evaluate algorithms from different domains.

\subsection{Challenges in Modeling Behavioral Changes}
We observe that for the three tools we studied, it is challenging for current models to predict users' future actions, particularly in open-ended tasks. These findings allude to a broader hypothesis that \emph{\textbf{current learning models may not be able to capture the evolution of users' exploration behavior in open-ended tasks}.} Subsequent studies are needed to verify the precise scope at which current learning models can capture users' exploration behavior over time.

For instance, in the goal-directed task (Section \ref{sec: forecache_results}), analysts use more interactions, and algorithms' performance improves with more training data. 
However, in the open-ended task (Section \ref{sec: imMens_results}), while we still observe analysts learning from new experiences, the performance of algorithms does not show the same improvement with increasing training data.
The issue mainly arises from the uncertainty of which action to choose next based on learned knowledge. While learning algorithms try to quantify this knowledge based on received rewards, their probabilistic approach fails when users decide to be exploratory. Because a user's prior knowledge, expertise in EVA, VES, and familiarity with the dataset may influence how to change her strategy, i.e., explore when the information need is not satisfied. 

Unlike goal-directed and focused tasks, open-ended tasks lack clear, pre-defined objectives. 
Therefore, users have extremely dynamic strategies. 
Given these evolving strategies for data exploration, users' learning algorithms may change and require complex algorithms to capture such nuances. 
As a potential solution to this challenge, we aim to try an ensemble of models for user behavior for complex data exploration tasks in our future work. 
In this way, the system can pivot gracefully when the current model fails. 
 
\subsection{Characteristics of User Learning}
\noindent
\textbf{Influence of Exploration Task Characteristics:} If we want to observe how users learn, we must place them in environments that challenge them to learn. 
For example, if users have \textbf{prior experience} with a task with similar \textbf{complexity} as the current task, it may not require as much effort.
This effect becomes apparent in the Tableau user study (Appendix \ref{sec: tableau}), where users complete three similar focused tasks before attempting the goal-directed task. 
Because of this prior experience, users often reuse attributes learned from focused tasks rather than exploring new ones. However, having limited prior experience still allows for the discovery of new data insights, as we observe in the Forecache user study.

%significantly hinder learning during the actual task. For example, consider users solving another goal-directed exploration task in \autoref{sec: forecache_results}. Before the task, users try \textit{1 training subtask} meant to familiarize users with the visualization tool. This kind of prior experience with low-complexity subtasks does not hinder users from still learning during the task.
These findings resonate with research by ~\citeauthor{csikszentmihalyi2014flow}, suggesting that observing learning in exploration tasks may require participants to balance prior experience with what is being asked of them, otherwise learning may not occur within the controlled lab environment. 

\noindent
\textbf{Learning Schemes:} 
Our statistical analysis of user behavior reveals differences in exploration strategies across tasks. 
However, our tested algorithms can satisfactorily model the evolution of users' exploration behavior for the \textbf{average user}. 
In the ForeCache user study (Section \ref{sec: forecache_results}), Reinforce, Actor-Critic, and Greedy sufficiently capture user learning, and even among these three, Reinforce is the top learning scheme for $70\%$ of the participants. 
Interestingly, {\bf simpler algorithms} outperform complex ones in modeling user exploration. 
For instance, Reinforce outperforms the more complex Actor-Critic in ForeCache's goal-directed task, while the Greedy approach closely matches the best-performing methods in the imMens and Tableau studies.
These findings endorse the preference of cognitive psychology and neuroscience researchers for simple algorithms as better candidates for modeling shifts in users' strategies~\cite{bennett2021value}.

%\leilani{This is a very nice result!}  
\subsection{Creating Versus Reusing Study Data}
The diversity of our selected studies from the scholarly works of visualization and HCI provides an opportunity to observe, formalize, and model diverse users and their exploration behaviors. 
Further, these works cover a large gamut of data exploration, enabling thorough testing of our hypotheses across diverse scenarios.
Additionally, designing a new user study poses significant challenges. First, recruiting a diverse user base is time-consuming. Second, devising tasks that span a wide range of open-endedness requires extensive research and domain expertise. Third, determining the appropriate visualization tools and level of detail to capture from user interactions adds to the layer of complexity~\cite{gathani2022grammar}.
Finally, as ~\citeauthor {zgraggen2018investigating} observe, users lack the awareness or desire to explain everything they learn during EVA. 
Thus, conducting a new user study might not be more beneficial to analyze users' exploration behavior than pursuing established studies.

\appendix
\section{Appendix}
\subsection{Markov Decision Process (MDP):\nopunct}
\label{subsection: mdp}
MDP is a framework for modeling decision-making in which an agent (user) learns to achieve its goals through repeated interactions with an environment (interface). In each interaction, the agent takes action based on its current state (exploration stage), and the environment (interface) responds with a reward (feedback). The agent uses these rewards to update its behavior and make better decisions in the future. The main objective of solving an MDP is to find an optimal \textbf{policy} ($\pi$), which is a function that maps each state to an action that maximizes the expected future reward.

\subsection{ForeCache User Study Algorithm Details}
\label{appendix: forecache_algorithms}
\noindent
{\bf Win-Stay-Lose-Shift (WSLS):} is a popular heuristic to model human learning in games, offering an alternative to randomization in bandit problems~\cite{tamura2015win}. This method repeats a successful strategy until it no longer yields rewards, then switches to other strategies with equal probabilities.

\noindent
{\bf Greedy:} requires the user to make decisions based on her previous experience. She decides upon an action that has yielded her the highest reward thus far \cite{sutton1998}.

\noindent
{\bf $\epsilon$-Greedy:} balances exploration-exploitation trade-off by choosing either a random action with a small probability ($\epsilon$) or the action with the highest estimated reward with probability (1 - $\epsilon$) \cite{zhang2013forgetful}.
\noindent
\\
\underline{Reinforcement learning (RL) algorithms selection criteria:}
In RL, \textit{value function (vf)} represents the expected future reward an agent can achieve by starting from a specific state and following a given policy\cite{sutton1998}. RL algorithms estimate and improve this {\it vf} through trial and error. Specifically, algorithms that learn the {\it vf} and take actions depending on this function are \textit{value-based algorithms}. These algorithms have been used to explain workings of the human brain \cite{glimcher2011dopamine}, and choice behavior~\cite{niv2009reinforcement,niv2012TD,daw2011modelbased}. Qlearning(QLearn) and State-Action-Reward-State-Action (SARSA) are two popular value-based algorithms.

%value-free/policy gradient definition, use of such algorithms
Alternatively, other cognitive psychology researchers~\cite{bennett2021value} advocate for algorithms that directly learn a policy that can select actions without a {\it vf}. A {\it vf} may still be used to learn the parameters defining a policy but is not required for action selection~\cite{sutton1998}. 
These algorithms are called \textit{value-free or policy-gradient }. Researchers in~\cite{bennett2021value} also suggest algorithms combining value-based and policy gradients that have shown promising results in modeling activities in neural structures~\cite{joel2002actorcritic} and explaining human behavior~\cite{bennett2022model}. Reinforce and ActorCritic are two popular policy-gradient based algorithms.

\noindent
\textbf{QLearning (Qlearn):} iteratively updates a value function called the Q-value. Qlearn learns the optimal policy through trial and error, following a different policy ($\epsilon$-greedy policy) during training. The goal of QLearn is to learn a policy that maximizes the expected reward in an environment~\cite{watkins1992q} based on the Q-value update rule:
\[ 
Q(s_t, a_t) \leftarrow Q(s_t, a_t) + \alpha[r_t + \gamma \max_{a} Q(s_{t+1}, a) - Q(s_t, a_t)]
\]

Where $Q(s_t, a_t)$ is the value of taking action $a$ in state $s$ at time $t$. $\alpha$ is the learning rate and controls for the degree of Q-value (Q) update, $r_t$ is the reward received at time $t$, $\gamma$ is the discount factor to give more weight to $r_t$ than future rewards, and $s_{t+1}$ is the next state. The last hyper-parameter in this algorithm is $\epsilon$ for $\epsilon$-greedy.

\noindent
{\bf State Action Reward State Action (SARSA):}
is value-based like QLearn~\cite{rummerysarsa1994}. But unlike Qlearn, which updates its Q using the action that yields \textbf{maximum} Q-value in the next state, SARSA updates Q by following the action based on the $\epsilon$-greedy policy ($\mathbf{a_{t+1}}$):
\[ 
Q(s_t, a_t) \leftarrow Q(s_t, a_t) + \alpha[r_t + \gamma Q(s_{t+1}, \mathbf{a_{t+1}}) - Q(s_t, a_t)]
\label{eqn: sarsa}
\]

SARSA has the same hyperparameters as QLearn: $\gamma$, $\alpha$, $\epsilon$

\noindent 
{\bf Reinforce:}
is the simplest policy-gradient method~\cite{sutton1998}. It directly improves the policy based on the observed rewards without any value function~\cite{williams1992reinforce}. A set of parameters define the policy. These parameters are represented by a neural network and improved by following the gradient of the expected reward with respect to the parameters. Reinforce uses $\gamma$ and $\alpha$ as hyperparameters.

\noindent 
{\bf Actor-Critic:}
extends Reinforce by improving the policy through learning a value function in parallel. It combines value-based (critic) methods with a policy-gradient side (actor)~\cite{konda2003actor}. A neural network acts as a function approximator to learn the policy and value parameters. This algorithm has the same hyperparameters as Reinforce.

\subsection{imMens user study Statistical Test}
\label{subsec: imMens_wilcox}
Table \ref{tab:immens_wilcox} presents the outcomes of a statistical test conducted on two halves (initial and later) of participants' exploration sessions. All p-values exceed $0.05$, indicating no significant differences observed between the initial and later exploration phases for any visualizations.

\begin{table}[H]
\begin{tabular}{|c|c|c|c|}
\hline
\begin{tabular}[c]{@{}c@{}}Explored\\ Dataset\end{tabular} &
  Visualization &
  P-value &
  Statistic \\ \hline
\multirow{4}{*}{\begin{tabular}[c]{@{}c@{}}Flight \\ Performance\end{tabular}} &
  \begin{tabular}[c]{@{}c@{}}Arrival Delay\\ vs Departure\\ Delay\end{tabular} &
  0.055 &
  4.0 \\ \cline{2-4} 
 & Carrier    & 0.11 & 6.0  \\ \cline{2-4} 
 & Year       & 0.46 & 12.0 \\ \cline{2-4} 
 & Month      & 1.0  & 14.0 \\ \hline
\multirow{5}{*}{\begin{tabular}[c]{@{}c@{}}Travelers'\\ Check-in\end{tabular}} &
  \begin{tabular}[c]{@{}c@{}}Geographical\\ Map\end{tabular} &
  0.11 &
  6.0 \\ \cline{2-4} 
 & Month      & 0.31 & 10.0 \\ \cline{2-4} 
 & Day        & 0.13 & 5.0  \\ \cline{2-4} 
 & Year       & 0.74 & 15.0 \\ \cline{2-4} 
 & Travellers & 0.84 & 16.0 \\ \hline
\end{tabular}
\caption{Wilcoxon-Signed-Rank Test Results }
\label{tab:immens_wilcox}
\end{table}

\subsection{imMens user study Algorithm details:}
\label{appendix: imMens_algorithm}
\noindent
\textbf{Mortal-arm bandit:} 
Assume we have a K-arm bandit machine with unknown stochastic rewards.
In each round, an agent pulls one arm and receives the reward.
Mortal-arm bandit~\cite{chakrabarti2008mortal} (MoAB) introduces the concept of mortality, where each arm $(i)$ has a finite lifetime ($L_i$).
Once an arm's lifetime ends, it is removed from the K-arm bandit and a new arm seamlessly replaces it, ensuring the bandit maintains a constant size of $K$ arms.

The authors propose two mortality implementations: \textit{budgeted death}, where each arm dies after $L_i$ selections drawn from a geometric distribution with an expected budget of $L$, and \textit{timed death}, where each arm dies with probability $p$ and have a lifetime of $L=1/p$. 
% Originally used in the online-advertising setting, 
The algorithm's objective is to maximize the expected total reward by selecting an optimal sequence of arms to pull. 
 
In the state-oblivious algorithm for MoAB~\cite{chakrabarti2008mortal},   
instead of pulling each arm once to estimate the payoff, each arm is pulled $n$ times and abandoned if deemed unfavorable. 
The objective function of MoAB minimize regret, which is the difference between the expected payoff of the best alive arm and the payoff obtained by MoAB.

In short, MoAB considers that each visualization has a lifetime, after which the user will not use it again. Maybe the user has extracted all the information she wanted from that visualization. 
Additionally, we have the flexibility to introduce the killed visualization as users' future exploration choice.
\\
\noindent
\textbf{Contextual multi-arm bandit:}
In our scenario, we have a set of $K$ visualizations. At each time step ($t$), the user selects a specific visualization (action in this case) to interact with, resulting in a reward $r_t$. In the contextual bandit framework, an agent, guided by observed context ($c_t$), chooses a visualization ($v_t$) as an action and receives rewards solely for the chosen action. Context, i.e., additional side information, assists the agent in maximizing the reward function. Our implementation uses the user's raw interactions (like pan, brush, range select, and zoom) as context, assuming potential latent correlations with visualization selection. For instance, users might frequently perform pan operations on geographical heat maps rather than other raw interactions. Contextual multi-arm bandit frameworks are particularly beneficial in non-stationary environments—dynamically changing scenarios \cite{li2010contextual}—with small action spaces.
% \section{Focused \& goal-directed open-ended}
%\vspace{-1mm}
\section{Tableau user study}
\label{sec: tableau}
In this user study by Battle et al.~\cite{battle2019characterizing}, participants perform a series of focused and open-ended exploration tasks (\textit{ordered by their open-endedness}) based on a particular dataset (table \ref{tab:tableau-table}). 
These tasks encapsulate nearly the full spectrum of task complexity and open-endedness and provide the opportunity to investigate the nature of user learning in a wide range of exploration scenarios.

% It provides us the opportunity to analyze user learning within each of these tasks as well as collectively.
%So, we are going to discuss this user study first. 
\begin{comment}
We prioritize the analysis of this user study as it allows us to examine learning behavior across tasks with different open-ended characteristics (\autoref{tab:user_study_high_level}) on the same dataset.
The completion of focused tasks before the goal-directed open-ended task presents an opportunity to investigate how the former influenced the latter.
\end{comment}

% \leilani{Perhaps the ``overview of exploration task'' should be listed under the Methodology in 2.2.}

\vspace{-3mm}
\subsection{Overview of Exploration Task}
% \leilani{We need to explain why we are providing this overview. It looks like we're explaining the facets of the dataset that we leveraged for modeling? (I'm not sure, please confirm)}
% \subsubsection{Dataset: Tableau interaction log}
\subsubsection{Analysis Tasks:}
27 Participants who use Tableau \cite{Tableau}  regularly for academic or professional purposes are selected to complete a series of analysis tasks with varying requirements (e.g., table \ref{tab:tableau-table}).
Their expertise varied widely in Tableau and data analysis experience, from just learning Tableau to seasoned veteran analysts to Tableau power users.

The datasets users explore are: (a) Weather station reports encompassing weather metrics and phenomena (35 columns, 56.2M rows), (b) US domestic flight performance data (31 columns, 34.5M rows), and (c) Aircraft striking wildlife reports, including contextual details (94 columns, 173K rows) \cite{battle2019characterizing}.

\begin{table*}[!t]
\small{%
\begin{tabular}{|l|l|l|}
\hline
              & Task & \multicolumn{1}{c|}{Task Description}                                                                                                          \\ \hline
\multirow{3}{*}{\rotatebox[origin=c]{90}{\centering Focused}} &
  T1 &
  \begin{tabular}[c]{@{}l@{}}Consider the following weather measurements: Heavy Fog{[}Heavy Fog{]}, Mist {[}Mist{]}, Drizzle {[}Drizzle{]}, and Ground \\ Fog {[}Ground Fog{]}. Which measurements have more data?\end{tabular} \\ \cline{2-3} 
 &
  T2 &
  \begin{tabular}[c]{@{}l@{}}How have maximum temperatures {[}T Max{]} and  minimum temperatures {[}T Min{]} changed over the\\ duration of the dataset (i.e., over the {[}Date{]} column)?\end{tabular} \\ \cline{2-3} 
              & T3      & \begin{tabular}[c]{@{}l@{}}How do wind measurements {[}High Winds{]} compare for the northeast and southwest regions of the US?\end{tabular} \\ \hline
{\rotatebox[origin=c]{90}{\parbox[c]{1.3cm}{\centering Goal\\Directed}}} & T4      & \begin{tabular}[c]{@{}l@{}}What weather predictions would you make for  February 14th 2018 in Seattle, and why?\end{tabular}                 \\ \hline
\end{tabular}%
}
\caption{Analysis tasks for Weather dataset}
\vspace{-4mm}
\label{tab:tableau-table}
\end{table*}

\subsubsection{Task Characteristics:}
\label{sec:tableau_task_charactertistics}
In this user study, the \textit{focused tasks} contain \emph{explicit hints} on which columns (inside square brackets in Table \ref{tab:tableau-table}) to exploit for task completion. 
It allows us to investigate user learning, where the scope of exploration is limited, as users only need to locate the hinted data area. 
Also, we examine how learning affects tasks with different \textit{open-ended characteristics} and varying task requirements (e.g., data quality assessment, evaluation of relationships between variables, causality, and prediction analysis).
Furthermore, the lack of \textit{prior knowledge} about the dataset, and the use of Tableau (with a \textit{complex interface}),  enriches our research of user learning in various exploration scenarios.
% compared to the other systems studied in this paper.

\subsubsection{Interface and Interaction Log:}
\begin{figure}[!htbp]
\centering
\includegraphics[scale=0.33]{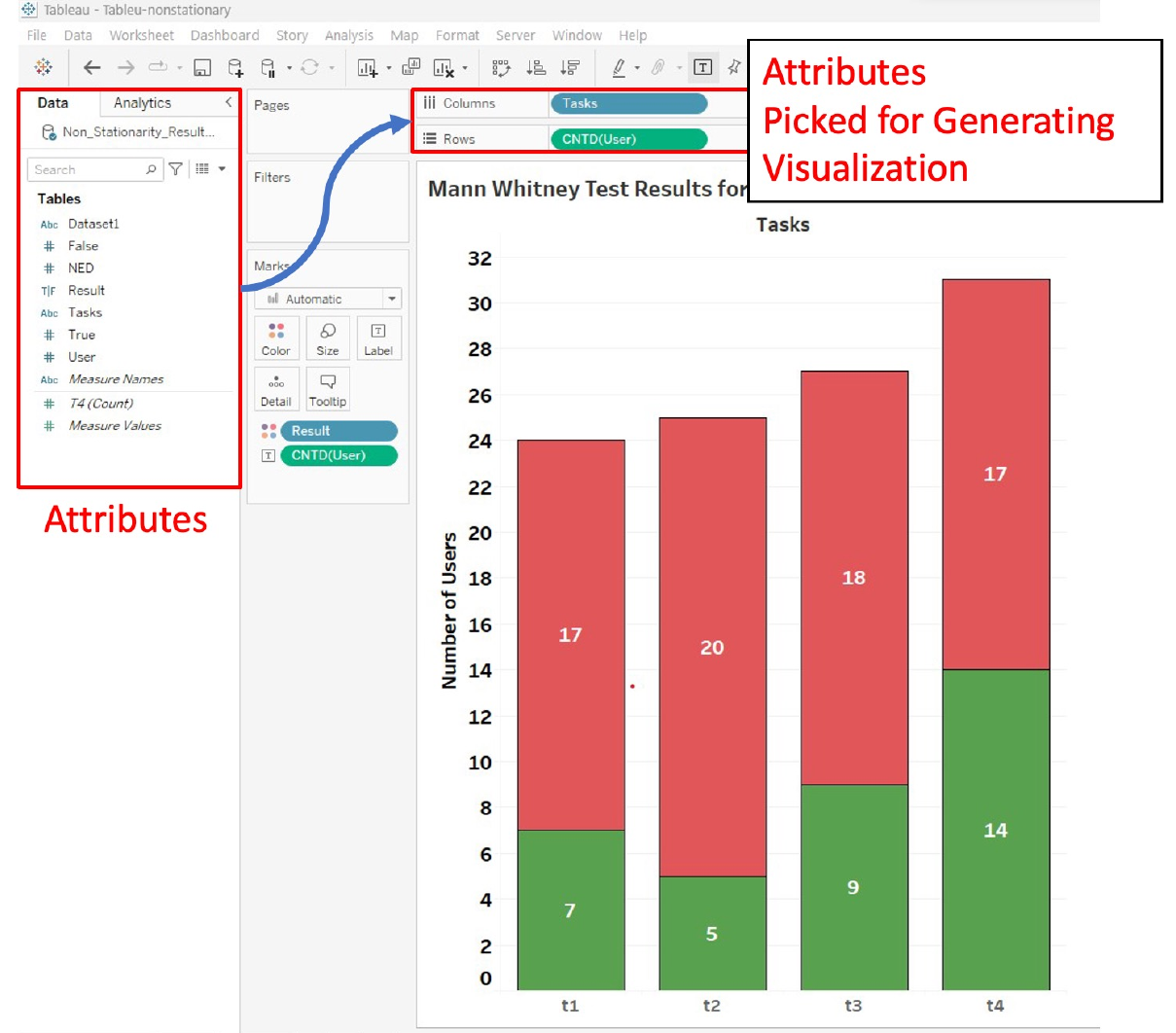}
\vspace{-3mm}
\caption{Analyzing attributes using Tableau interface}
\label{fig: tableau_interface}
\vspace{-7mm}
\end{figure}

Tableau presents \textbf{\textit{attributes}} (columns of a dataset) for user exploration (shown in Figure \ref{fig: tableau_interface}).
Users can choose which attribute to analyze and add to the Tableau Worksheet.
Tableau then suggests visualizations, from which users can choose based on their interpretability. 
For our research, we analyze participants' interactions with the Tableau interface as recorded by Battle et al.~\cite{battle2019characterizing}, who completed all tasks.
% Participants add desired \textbf{\textit{attributes}} (columns of a dataset) to the Tableau interface and generate easy-to-interpret visualizations.
% It eases and accelerates learning and insight discovery in the focused data area.
The interaction log contains information on users' exploration activities, such as their choice of attributes, visualization, and time spent in each interaction. 
Participants provide feedback for the open-ended tasks, highlighting meaningful insights in their problem-solving approach. 
\vspace{-3mm}
\vspace{2mm}
\subsubsection{Attributes Consolidation:}
% \leilani{I don't understand the first two sentences.}
In focused and goal-directed tasks, users start exploration with a clear understanding of the tasks' requirements.
So, users may need fewer interactions than open-ended tasks to extract desired information.
In this study, the given datasets are large and have many attributes but relatively few interactions.
For example, the wildlife strikes dataset contains 94 attributes but has an average of 24 interactions across tasks.
It poses a significant challenge in empirically analyzing and modeling user learning.

Thus, we reduce the search space for our learning algorithms by identifying attributes from similar domains and consolidating them into high-level attributes.
For example, we observe the attributes \textit{t\_max}, \textit{t\_min}, \textit{t\_minf} and \textit{t\_maxf} (from T2 in table \ref{tab:tableau-table}) all represent maximum and minimum temperature in C$^{\circ}$ and F$^{\circ}$ respectively. 
So, we consolidate them into high-level attributes: temp\_max and temp\_min.
We are careful in this consolidation process so that it does not change users' intentions behind each interaction, which can affect our experiments on user learning. 
% \leilani{I would say grouping or consolidation rather than aggregation, so the term is not overloaded, since aggregation can also describe how table rows are manipulated.}
\vspace{-3mm}
\subsection{Formalizing User Learning Problem}
\label{sec: tableau_user_learning}
\subsubsection{User Activities:}
\label{sec: tableau_user_activities}
One approach to analyzing how users' exploration behavior evolves is to examine the low-level Tableau interactions performed at each step. 
However, Tableau contains many interaction paths that ultimately lead to the same underlying data manipulations~\cite{battle2019characterizing}.
%But users complete the given tasks with limited interactions, utilizing a large action space from which we may not capture existing strategy changes. 
Thus, instead of focusing on how users navigate Tableau's massive action space, we instead analyze which attributes (i.e., data areas) users select to achieve their goals.
% To overcome this issue, we consider \textit{identifying and analyzing the necessary attributes with essential information as the user's high-level goal}.

\textbf{In each interaction, the users select a subset of attributes from the dataset} to generate visualizations.
They use information from these visualizations to make the following decisions, which \textit{influences selection of future action}, i.e., \textit{exploration strategy}:
(a) spend more time understanding the current data area, or (b) switch focus to other data areas for additional information or new findings. 

\noindent
\underline{\textit{Actions: }}
In each interaction, the user can use the actions, \textbf{Keep, Add, Drop, and Reset}, to modify the specific set of attributes she currently has on the Tableau worksheet for analysis.
% Let us discuss what each of them does:
% (a) 
{\bf Keep} reuses the same set of attributes from the previous time step.
% (b) 
{\bf Add} incorporates one or more attributes from the dataset into the current set.
% (c) 
{\bf Drop} removes one or more attributes from the current set.
% (d) 
\textbf{Reset} removes all attributes from the worksheet and starts anew.

\noindent
\underline{\textit{Reward:}}
The user analyzes the Tableau visualization generated for her selected set of attributes. 
The \textit{information in this visualization acts as a reward} through which she can measure the relevance of the current set of attributes to complete the analysis task.
Based on the reward, the user decides her next set of actions.

In our experiment, we quantify the relevance of users' selected set of attributes.
First, we identify the \textit{attributes necessary (i.e., ground truth)} for task completion based on users' feedback and by analyzing the task requirements, dataset, and users' interactions.
% First, we analyze users' feedback and task requirements to identify \textit{necessary} (i.e., ground truth) attributes for task completion.
Then, for each interaction, we use the \textit{size of the intersection between users' current selected attributes and the necessary ones for the tasks as a reward.}
The utilization of necessary attributes is crucial for completing analysis tasks effectively. Increased usage of these attributes to create visualizations leads to more relevant information.
% We identify the essential attributes required for completing each analysis task, deeming them high-quality.
% The effectiveness of a selected set of attributes = \#of high-quality attributes in it. 

\subsubsection{Learning Problem:}
\label{learning_tableau}
A user's exploration strategy determines which action ($a_t$) to pick in interaction at time step $t$, spanning from the initial interaction at $t = 1$ to the last at $t = T$.
At $t$, action choice $a_t$ depends on the user's learned decision to keep, modify, or reset the current attribute set based on the received reward $r_t$.
The exploration strategy outlines the decision-making process to optimize the objective function $f(a_t | (a_{\hat{t}}, r_{\hat{t}})) = \sum_{t}^{T} \mathbb{E}[Reward]$.
Here, for time t, $f$ quantifies the expected information gain of action $a_t$ based on past actions $a_{\hat{t}}$ and corresponding rewards $r_{\hat{t}}$ (where $\hat{t} = 1, ..., t - 1)$. 
$f$ can be optimized by maximizing rewards. 
\textit{The learning problem involves optimizing $f$ online to find the optimal strategy that will lead to the desired information.}
Using the optimal strategy, the user will pick an action that will influence the selection of ground truth attributes with the necessary information (yielding high rewards) for task completion.  
\vspace{-3mm}
\subsection{Evaluated Learning Algorithms}
\label{sec: tableau_methods}
Here we discuss the human learning algorithms used in this user study.
We use these algorithms' objective functions to model users' exploration behavior during EVA tasks.
% for what humans optimize to complete IDE tasks.

\subsubsection{Random Strategy:}
In this approach, the agent always picks an action \textit{uniformly at random} from the available choices. Action choice is made irrespective of the rewards received or consideration for potential outcomes. This strategy serves as the \textit{baseline}.

\subsubsection{Heuristics:}
Greedy and Win-Stay Lose-Shift may model user learning using prior experiences with relatively simple heuristics. 

% These algorithm's action choice derives from applying simple heuristics based on past information. 
\noindent 
% {\bf Greedy} requires the user to make decisions based on her previous 
{\bf Greedy} requires the user to pick an action for immediate success based on her previous experience. 
She chooses the action that has yielded her the highest reward thus far, hoping that it will increase her cumulative gains~\cite{sutton1998}.

\noindent
\textbf{Win-Stay Lose-Shift} is a popular heuristic to model human learning in games, offering an alternative to randomization in bandit problems~\cite{tamura2015win}. It repeats a successful action until it no longer yields rewards, then switches to other actions with equal probabilities.

The simplicity of these approaches does not ensure globally optimal solutions. The primary objective of these approaches is to maximize the cumulative reward in the sequential decision-making process.
%But to achieve that, they rely on maximizing immediate gains based on past actions rather than the potential implications of being stuck with suboptimal solutions.
But to achieve that, they rely solely on maximizing immediate rewards based on past actions, making them vulnerable to being stuck with suboptimal solutions.

\subsubsection{Learning Algorithms from Game Theory:}
Bush \& Mosteller \cite{bush1953stochastic} and Roth \& Erev \cite{erev1995need, young2004strategic} have been popular in modeling human learning in games. 
Recent research shows their success in modeling human learning in information searching \cite{mccamish2018data, cen2013reinforcement}.   

\noindent
\textbf{Bush and Mosteller} updates the probability of using an action at time step $t$, $a(t)$ by an amount proportional to the received reward $r(t)$ for using this action and its current probability~\cite{bush1953stochastic}. If the user uses action $a_i \in {1, ..., n}$ for a(t) then the model updates the probability distribution of the strategies $P_i$ as follows:
% \vspace{-3mm}
\begin{equation}
    P_i(t + 1) =  \left\{\begin{matrix}
        P_i (t) + \alpha \times (1 - P_i (t) ) \&\& a(t) = a_i, r(t) \geq 0 \\ 
        P_i (t) - \beta \times P_i (t) \&\& a(t) = a_i, r(t) < 0 \\
        \end{matrix} \right .
\end{equation}

\begin{equation}
    P_i(t + 1) =  \left\{\begin{matrix}
        P_i (t) - \alpha \times P_i (t) \&\& a(t) \neq a_i, r(t) \geq 0 \\ 
        P_i (t) + \beta \times (1 - P_i (t) ) \&\& a(t) \neq a_i, r(t) < 0 \\
        \end{matrix} \right .
\end{equation}
$\alpha \in [0, 1]$ and $\beta \in [0, 1]$ are the only hyper-parameters, where $\alpha$ determines the weighting for non-negative rewards, while $\beta$ regulates negative rewards.

\begin{table*}[!h]
\begin{tabular}{|l|l|c|c|c|c|c|c|c|c|}
\hline
Datasets &
  Tasks &
  \begin{tabular}[c]{@{}c@{}}Random\\ Strategy\end{tabular} &
  \begin{tabular}[c]{@{}c@{}}Win-Stay\\  Lose-Shift\end{tabular} &
  \multicolumn{1}{l|}{\begin{tabular}[c]{@{}l@{}}Greedy \\ algorithm\end{tabular}} &
  Roth \& Erev &
  \multicolumn{1}{l|}{\begin{tabular}[c]{@{}l@{}}Bush \& \\ Mosteller\end{tabular}} &
  \multicolumn{1}{l|}{$\epsilon$-greedy} &
  \begin{tabular}[c]{@{}c@{}}Adaptive \\ $\epsilon-$Greedy\end{tabular} &
  \begin{tabular}[c]{@{}c@{}}Combinatorial\\ Bandit\end{tabular} \\ \hline
\multirow{2}{*}{\begin{tabular}[c]{@{}l@{}}Wildlife\\ Strikes\end{tabular}}   & T3 & 0.17 & 0.19 & \textbf{0.78} & 0.68 & 0.71 & \textbf{0.73} & \textbf{0.73} & 0.43 \\ \cline{2-10} 
                                                                              & T4 & 0.17 & 0.16 & \textbf{0.65} & 0.59 & 0.56 & \textbf{0.65} & \textbf{0.65} & 0.38 \\ \hline
\multirow{2}{*}{Weather}                                                      & T3 & 0.14 & 0.15 & \textbf{0.82} & 0.69 & 0.67 & \textbf{0.83} & \textbf{0.84} & 0.45 \\ \cline{2-10} 
                                                                              & T4 & 0.13 & 0.13 & \textbf{0.77} & 0.72 & 0.70 & \textbf{0.77} & \textbf{0.78} & 0.32 \\ \hline
\multirow{2}{*}{\begin{tabular}[c]{@{}l@{}}Flight\\ Performance\end{tabular}} & T3 & 0.12 & 0.13 & \textbf{0.64} & 0.50 & 0.56 & \textbf{0.65} & \textbf{0.62} & 0.40 \\ \cline{2-10} 
                                                                              & T4 & 0.13 & 0.14 & \textbf{0.65} & 0.57 & 0.54 & \textbf{0.66} & \textbf{0.65} & 0.30 \\ \hline
\end{tabular}
\caption{Recall (K=3) value of discussed learning algorithms in the action prediction task.}
\label{tab:tableau_recall}
\end{table*}

\noindent
\textbf{Roth and Erev} reinforces action probabilities based on the rewards received after each action. 
Forgetting parameter, $\sigma \in [0,1]$, controls the degree to which past outcomes influence future decisions \cite{erev1995need, young2004strategic}.
A matrix, $S(t)$, maintains the accumulated reward for using different actions over time, $t$. 
Its cell ($i, t+1$) is updated after an action $a_i$ is performed using, $S_i(t+1) = S_i(t) \times (1 - \sigma) + r$. Given the users have $n$ actions to pick from, the probability of performing action $a_j$ after time $t$ is: $P_j(t+1) = S_j(t+1)\} / {\sum_{j'}^{n} S_{j'}(t+1)}$ 
   
% \begin{equation}
% \label{eqn: basicroth1}
%     S_i(t + 1) =  \left\{\begin{matrix}
%         S_i(t) \times (1 - \sigma) + r && i = a(t)\\ 
%         S_i(t) && i \neq a(t)\\
%         \end{matrix} \right .
% \end{equation}
% \begin{equation}
% \label{eqn: basicroth2}
%     P_j(t + 1) = \frac{S_j(t + 1)}{\sum_{j'}^{n} S_{j'}(t + 1)}
% \end{equation}

It is important to note that these approaches do not have a fixed or standardized objective function. Intuitively, they try to find the action(s) that returns the most payoff in the future.

\subsubsection{$\epsilon$-Greedy Based Algorithms:}
% Users make a learned choice whether to exploit the current attributes or explore different ones for answers. 
% This decision-making process is similar to the 
$\epsilon$-Greedy and Adaptive $\epsilon$-greedy algorithms aim to maximize the overall cumulative reward by balancing the exploration-exploitation trade-off.  
% for modeling user learning. 

\noindent
{\bf $\epsilon$-Greedy} balances exploration-exploitation trade-off by choosing either a random action with a small probability ($\epsilon$) or the action with the highest estimated reward with probability (1 - $\epsilon$) \cite{zhang2013forgetful}.

% \noindent
% {\bf $\epsilon$-Greedy with $\epsilon$ decay:} In this version of the $\epsilon$-Greedy, $\epsilon$ gradually decreases over time\cite{sutton1998}. This gives  the algorithm additional support to explore more in the early stages and exploit later on. 

\noindent
{\bf Adaptive $\epsilon$-Greedy} extends $\epsilon-$Greedy by changing the value of $\epsilon$ during learning using two hyper-parameters, $l$ and $f$. $l$ keeps track of how many times to run exploration before performing the adaptive action that changes the value of $\epsilon$. $f$ is for regularizing the change in average accumulated reward $\Delta$, before and after the previous $\epsilon$ changes. 
The new value of $\epsilon$ is created using $sigmoid (\Delta)$, $\Delta =$ $(reward_{current-\epsilon} -$ $reward_{previous-\epsilon}) \times f$ \cite{dos2017adaptive}.

\subsubsection{Combinatorial Multi-arm Bandit (CMAB):}
To aid in understanding, we will use the terminologies from our learning problem rather than the generic CMAB terms, e.g., \textit{actions instead of arms}.

Let us consider a scenario where there are $N$ actions, and at each step, a user chooses a subset of actions and receives rewards. 
We denote the set of all possible combinations or subsets of actions as $A$ (size = $n$), represented by variable $A = \{A_1, ..., A_n\}$.
% Each variable $X_i$ consists of $K_i$ different actions $X_i = \{v^{1}_i, ... , v^{K_i}_i\}$.
The reward $R: \{\alpha_1 \times ... \times \alpha_{K_i}\} \rightarrow \mathbb{R}$ depends on the reward of the $K_i$ actions  in the picked subset $(A_i)$. 
The goal of the problem is to find a vector $V \subseteq \{0, 1\}^n$ representing the subset combination of actions that minimizes the regret, $\rho(\pi, t) = T \mu^* - \sum_{t = 1}^{T} R(a^{t}_1, ..., a^{t}_n)$
\cite{ontanon2013combinatorial, ontanon2017combinatorial}. 
The regret equation is for policy $\pi$, where $a^{t}_1, ..., a^{t}_n$ are the actions selected at time step T. 
% \vspace{-3mm}
% \begin{equation}
%     \rho(\pi, t) = T \mu^* - \sum_{t = 1}^{T} R(x^{t}_1, ..., x^{t}_n)
%     \label{eqn: combinatorial_regret}
%     \vspace{-1mm}
% \end{equation}
Here, $\mu^{*} = \mathbb{E}(R(v^{*}_1, ..., v^{*}_n))$ is the maximum expected reward gained by following the optimal policy $\pi^{*}$.

\noindent
% {\bf Combinatorial Bandit:} For this implementation, we
{\bf Used approach:} To solve the CMAB problem, we use the Vowpal Wabbit library \cite{VowpalWabbit, swaminathan2017off}, whose CMAB implementation is heavily influenced by Bianchi and Lugosi \cite{cesa2012combinatorial}.
In this approach, the algorithm picks an action subset $K_t \in A$ for each time step, $t = 1,2,..$ based on the distribution, $p_{t-1} = (1 - \gamma)q_{t-1} + \gamma\mu$. 
% To solve the CMAB problem, we use the proposed algorithm in \cite{cesa2012combinatorial}.
% According to the proposed algorithm in \cite{cesa2012combinatorial}, given all possible subsets of arms (action set), $S \subseteq \{0, 1\}^d$, the algorithm picks an action $K_t$ for each time step, $t = 1,2,..$ based on the distribution, $p_{t-1} = (1 - \gamma)q_{t-1} + \gamma\mu$. 
Where $q_t(k) = \overline{w}_t(k) / \overline{W}_t$ and $\overline{W_t}$ corresponds to a weight vector that keeps track of the cumulative pseudo-loss. 
Upon predicting $K_t$, the algorithm observes cost $l_t(K_t)$. 
This cost updates a vector of pseudo-loss $\Tilde{l_t}$, which consequently is used to update the $q_t(k)$.
Prior distribution $\mu$ and mixing coefficient $\gamma$ are the hyper-parameters.

\subsection{Performance Evaluation}
\label{sec:tableu_empirical_evaluation}
\subsubsection{Evaluation Procedure:}
\label{sec:tableu_evaluation_procedure}
To evaluate the discussed algorithms' (\ref{sec: tableau_methods}) performance in modeling user learning, we use them to predict \textbf{What \textit{action(s)} a user will use in her next interaction.}
How many attributes a user may add or drop in a future interaction is extremely difficult to predict and beyond the scope of this paper.
To simplify our evaluation process, we employ algorithms to predict three actions for the next interaction, following a recall $(k = 3)$ assessment.
It is decided based on the average number of actions by the users in this user study.

%Learning algorithms' true adaptability to users' exploration behavior is best tested on non-stationary cases. 
Note that since users rarely need to and lack evidence of learning within $T1$ \& $T2$, we exclude these tasks from evaluation.

\subsubsection{Analyzing Performance of the Learning Models:}
\label{sec:tableu_perfromance_evaluation}
We train the hyperparameters using 20$\%$ of the data to ensure that our models are properly tuned.  
Results in Table \ref{tab:tableau_recall} show users mainly adopt the Greedy strategy. 
We also observe that users keep using the same set of attributes and exploit them for a long time when they yield high rewards.
The $\epsilon-$Greedy and adaptive $\epsilon-$ Greedy experiments support this observation, as we get the best results for $\epsilon = 0.001$. 
Such a low $\epsilon$ value indicates users rarely do any exploration and have a preference for exploitation based on past experiences.

Win-Stay Lose-Shift (WSLS) performs the worst as it has to pick future actions randomly at the start. 
However, even after categorizing the attributes, our action space is still large (e.g., 22 for weather) compared to the number of interactions.
Under these circumstances, it is difficult for the models to choose 3 attributes randomly and get them correct in a limited number of interactions.
Our Combinatorial bandit algorithm suffers from the same issue, although we fixed the subset size to 3 before generating all possible combinations. 

Roth \& Erev and the Bush \& Mosteller model are slower than the Greedy approaches in adapting to users' current information needs, contributing to subpar performance. 

\bibliography{aaai24}

\end{document}